\documentclass[12pt]{article}
\usepackage{latexsym}
\usepackage{amsmath,,calrsfs}
\usepackage{amsfonts}
\usepackage{amssymb}
\usepackage{amscd}
\usepackage{bbm}
\usepackage{fancybox}
\usepackage{cite}
\usepackage{amsmath,amsfonts,amsbsy}
\usepackage{pstricks,pst-node}
\usepackage{graphicx}
\usepackage{epsfig}
\usepackage{psfrag}
\usepackage{comment}
\usepackage{youngtab}
\usepackage{young}
\usepackage{slashed}

\psset{unit=1.3cm,linewidth=.5pt,radius=.2}  % controls the size of diagrams

\usepackage{multirow}                     % tables cells spanning multiple rows
\usepackage{float}                          % to force floaters to stay Here
\usepackage{lscape}                         % landscape laid out pages
\usepackage{bm}

\usepackage{color}

\definecolor{MyDarkBlue}{rgb}{0.15,0.15,0.45}
\definecolor{MyLightBlue}{rgb}{0.15,0.45,0.45}
\usepackage[linktocpage=true]{hyperref}
\hypersetup{
colorlinks=true,
citecolor=MyDarkBlue,
linkcolor=MyDarkBlue,
urlcolor=MyLightBlue,
}

\newcommand{\beq}{\begin{equation}}
\newcommand{\eeq}{\end{equation}}
\newcommand{\bi}{\begin{itemize}}
\newcommand{\ei}{\end{itemize}}
\newcommand{\bt}{\begin{tabular}}
\newcommand{\et}{\end{tabular}}
\newcommand{\bc}{\begin{center}}
\newcommand{\ec}{\end{center}}

\newcommand{\be}{\begin{equation}}
\newcommand{\ee}{\end{equation}}
\newcommand{\bea}{\begin{eqnarray}}
\newcommand{\eea}{\end{eqnarray}}
\newcommand{\ba}{\begin{array}}
\newcommand{\ea}{\end{array}}

\def\bbox{{\,\lower0.9pt\vbox{\hrule \hbox{\vrule height 0.2 cm
\hskip 0.2 cm \vrule height 0.2 cm}\hrule}\,}}
\newcommand{\dsl}{\pa \kern-0.5em /}

\makeatletter \@addtoreset{equation}{section} \makeatother

\def\slashchar#1{\setbox0=\hbox{$#1$}           % set a box for #1
   \dimen0=\wd0                                 % and get its size
   \setbox1=\hbox{/} \dimen1=\wd1               % get size of /
   \ifdim\dimen0>\dimen1                        % #1 is bigger
      \rlap{\hbox to \dimen0{\hfil/\hfil}}      % so center / in box
      #1                                        % and print #1
   \else                                        % / is bigger
      \rlap{\hbox to \dimen1{\hfil$#1$\hfil}}   % so center #1
      /                                         % and print /
   \fi}

\newcommand{\ga}{\gamma}
\newcommand{\Ga}{\Gamma}

\begin{document}

\begin{titlepage}%1
\begin{center}

\rightline{ UG-13-17}

\vskip 2.0cm

{\Large \bf  New Massive Supergravity and Auxiliary Fields}

\vskip 1,5cm

{\bf Eric A.~Bergshoeff\,${}^1$\,, Marija Kova\v{c}evi\'c\,${}^1$\,, Lorena Parra\,${}^{1,2}$\,,\\[.1truecm]
     Jan Rosseel\,${}^3$\,, Yihao Yin\,${}^1$} and {\bf Thomas Zojer\,${}^1$}\\

\vskip 25pt

{\em $^1$ \hskip -.1truecm  \em Centre for Theoretical Physics, University of Groningen, \\
                                Nijenborgh 4, 9747 AG Groningen, The Netherlands \vskip 5pt }

{email: {\tt e.a.bergshoeff@rug.nl, m.kovacevic@rug.nl, l.parra.rodriguez@rug.nl, y.yin@rug.nl, t.zojer@rug.nl}} \\

\vskip .5truecm

{\em $^2$ \hskip -.1truecm \em Instituto de Ciencias Nucleares, Universidad Nacional Aut\'onoma de M\'exico,\\
Apartado Postal 70--543, 04510 M\'exico, D.F., M\'exico }

\vskip .5truecm

% {\bf Eric A.~Bergshoeff\,${}^1$\,, Marija Kova\v{c}evi\'c\,${}^1$\,, Lorena Parra\,${}^1$\,,\\[.1truecm]
%      Jan Rosseel\,${}^2$\,, Yihao Yin\,${}^1$} and {\bf Thomas Zojer\,${}^1$}\\
% 
% 
% \vskip 25pt
% 
% {\em $^1$ \hskip -.1truecm  \em Centre for Theoretical Physics, University of Groningen, \\
%                                 Nijenborgh 4, 9747 AG Groningen, The Netherlands \vskip 5pt }
% 
% {email: {\tt e.a.bergshoeff@rug.nl, m.kovacevic@rug.nl, l.parra.rodriguez@rug.nl, y.yin@rug.nl, t.zojer@rug.nl}} \\
% 
% \vskip .5truecm

{\em $^3$ \hskip -.1truecm \em Institute for Theoretical Physics, Vienna University of Technology, \\
                               Wiedner Hauptstr.~8--10/136, A-1040 Vienna, Austria \vskip 5pt }

{email: {\tt rosseelj@hep.itp.tuwien.ac.at}} \\

\end{center}

\vskip 1cm

\begin{center} {\bf ABSTRACT}\\[3ex]
\end{center}

We construct a supersymmetric formulation of three-dimensional linearized New Massive Gravity without introducing higher derivatives. Instead, we
introduce supersymmetrically a set of bosonic and fermionic auxiliary fields which, upon elimination by their equations of
motion, introduce fourth-order derivative terms for the metric and third-order derivative terms for the gravitino. Our construction
requires an off-shell formulation of the three-dimensional supersymmetric massive Fierz--Pauli theory. We discuss the non-linear
extension of our results.

\end{titlepage}

\newpage
\setcounter{page}{1} \tableofcontents

% \newpage

%%%%%%%%%%%%%%%%%%%%%%%%%%%
\section{Introduction}\label{sec:intro}

New Massive Gravity (NMG) is a higher-derivative extension of three-dimensional (3D) Einstein--Hilbert gravity with a
particular set of terms quadratic in the 3D Ricci tensor and Ricci scalar \cite{Bergshoeff:2009hq}. The interest in the NMG model lies
in the fact that, although the theory contains higher derivatives, it nevertheless describes, unitarily, two massive
degrees of freedom of helicity +2 and $-$2. Furthermore, it has been shown that even at the non-linear level ghosts are absent
\cite{deRham:2011ca}. The 3D NMG model is an interesting laboratory to study the validity of the AdS/CFT correspondence in the
presence of higher derivatives. Its extension to 4D remains an open issue and has only been established so far at the
linearized level \cite{Bergshoeff:2012ud}.

For many  purposes, it is convenient to work with a formulation of the model
without higher derivatives, see, e.g.~\cite{Hohm:2010jc}. This can be achieved by introducing an auxiliary symmetric
tensor that couples to (the Einstein tensor of) the 3D metric tensor and has an explicit mass term \cite{Bergshoeff:2009hq}.
A supersymmetric version of NMG was constructed in \cite{Andringa:2009yc}. Besides the fourth-order-derivative terms of
the metric tensor this model also contains third-order-derivative terms involving the gravitino.

The purpose of this work is to
construct a reformulation of the supersymmetric NMG model (SNMG) {\sl without} higher derivatives. This requires that besides an
auxiliary symmetric  tensor, we introduce further auxiliary fermionic fields that effectively lower the number of derivatives of
the gravitino kinetic terms.

At the linearized level the NMG model decomposes into the sum of a massless spin-2 Einstein--Hilbert theory and a massive spin-2
Fierz--Pauli (FP) model \cite{Bergshoeff:2009hq}. In the supersymmetric case we therefore need a 3D massless and
a 3D massive spin-2 supermultiplet. We only consider the case of simple $\mathcal{N}=1$ supersymmetry. In this paper we will
explicitly construct the linearized, massive, off-shell spin-2 supermultiplet, paying particular attention to the auxiliary
field structure. We will obtain this massive spin-2 FP multiplet by starting from a 4D (linearized) massless spin-2
supermultiplet, performing a Kaluza--Klein (KK) reduction over a circle and projecting onto the first massive KK sector.
The final form of the 3D off-shell massive spin-2 supermultiplet is then obtained after a truncation and gauge-fixing a
few St\"uckelberg symmetries. Along with the construction of the massive, off-shell spin-2 multiplet, we will look in
detail at its massless limit. As is well-known, already in the bosonic case, this limit is non-trivial and should be
taken with care. Indeed, the massless limit of the massive spin-2 FP theory, coupled to a conserved energy-momentum
tensor, does not lead to linearized General Relativity, a result known as the van Dam--Veltman--Zakharov (vDVZ)
discontinuity \cite{vanDam:1970vg,Zakharov:1970cc}. Starting from the massive FP supermultiplet obtained, we will
explicitly illustrate how a supersymmetric version of this vDVZ discontinuity arises (see also \cite{Deser:1977ur}
for an earlier discussion).

Having constructed the off-shell massive spin-2 supermultiplet, it is rather straightforward to construct a linearized
version of SNMG without higher derivatives, by appropriately combining a massless and a massive spin-2 multiplet. This
theory contains three vector-spinors, whereas the higher-derivative version contains only one gravitino field. This is
due to the fact that the massive multiplet contains two gravitini, unlike the massless multiplet that only contains one.
The reason for this is that a massive gravitino describes a single helicity 3/2 state whereas $\mathcal{N}=1$ SNMG contains
two fermionic massive degrees of freedom of helicity +3/2 and $-$3/2. We will show that two of the vector-spinors are
actually auxiliary like the auxiliary symmetric tensor in the bosonic case. In particular, we will explicitly show how,
by eliminating the different bosonic and fermionic auxiliary fields, we re-obtain the linearized approximation of the
higher-derivative SNMG model given in \cite{Andringa:2009yc}. At the linearized level, we will
distinguish between two types of auxiliary fields: the ``trivial'' and ``non-trivial'' ones. The difference between them is that
only the elimination of the non-trivial auxiliary fields leads to higher derivatives in the action. The trivial ones are only
needed to obtain a supersymmetry algebra that closes off-shell.

The extension to the non-linear case, in the presence of both the trivial and non-trivial  auxiliary fields, is not obvious.
One way to see this, is by noting that our construction of the massive spin-2 multiplet is based upon a KK truncation which can
only be performed consistently at the linearized level. We consider the alternative option that first, at the linearized
level, one eliminates only the trivial auxiliary fields of the massive spin-2 supermultiplet but keeps all the other ones.
This implies that at the linearized level the supersymmetry algebra closes on-shell but that the action does not contain
higher derivatives. We will show that in principle the extension to the non-linear case in this situation is possible but
that the answer is not illuminating. This is in contrast to the higher-derivative formulation of SNMG where the contributions
to the bosonic terms of the single auxiliary scalar $S$ of the massless multiplet can be nicely interpreted as a torsion
contribution to the spin-connection \cite{Andringa:2009yc}.

This work is organized as follows. In section \ref{sec:sec2}  we show how the 3D supersymmetric Proca theory is obtained from
the KK reduction  of a 4D massless spin-1 Maxwell multiplet. This serves as an explanatory discussion for section
\ref{sec:sec3}, in which we extend this analysis to the spin-2
case and obtain the supersymmetric FP model. The corresponding supersymmetry algebra closes off-shell and contains 3 auxiliary
scalars and one auxiliary vector. In section \ref{sec:sec4} we use these results to construct a linearized version of SNMG
without higher derivatives. We explicitly show how, after eliminating all bosonic and fermionic auxiliary fields, the higher
derivatives of the metric and gravitino are introduced. In section \ref{sec:sec5} we discuss our attempts to extend our results
to the non-linear case. Our conclusions are presented in section \ref{sec:sec6}. There are two appendices. In appendix
\ref{app:multiplets} we summarize some properties of the off-shell massless multiplets that occur in this work. In appendix
\ref{app:fermions} we show, as a spin-off of the main discussion, how the trick that can be used to boost up the derivatives
in the FP model, see e.g.~\cite{Bergshoeff:2009tb}, can be extended to the fermionic case to boost up the number of derivatives
in a massive gravitino model.

\section{Supersymmetric Proca}\label{sec:sec2}

In this section we show how to obtain the 3D supersymmetric Proca theory from the KK reduction of an off-shell 4D $\mathcal{N}=1$
supersymmetric Maxwell theory and a subsequent truncation to the first massive KK sector. This is a warming-up exercise for the
spin-2 case which will be discussed in the next section.

\subsection{Kaluza--Klein reduction}

Our starting point is the 4D $\mathcal{N}=1$ supersymmetric Maxwell multiplet which consists of a vector ${\hat V}_{\hat\mu}$, a
4-component Majorana spinor ${\hat \psi}$ and a real auxiliary scalar ${\hat F}$. We indicate fields depending on the
4D coordinates and 4D indices with a hat. We do not indicate spinor indices. The supersymmetry rules,
with a constant 4-component Majorana spinor parameter $\epsilon$, and gauge transformation, with local parameter ${\hat \Lambda}$, of
these fields are given by
\begin{align}\begin{split}\label{v4rules}
 \delta \hat{V}_{{\hat\mu}} &= -\bar{\epsilon}\Ga_{\hat{\mu}}\hat{\psi} +\partial_{\hat{\mu}}\hat{\Lambda} \ ,\\[.1truecm]
 \delta \hat{\psi}  &= \frac18\Ga^{\hat{\mu}\hat{\nu}}\hat{F}_{\hat{\mu}\hat{\nu}} \epsilon +\frac14i\Ga_5\hat{F}\epsilon \ ,\\[.1truecm]
 \delta \hat{F}     &= i\bar{\epsilon}\Ga_5\Ga^{\hat{\mu}} \partial_{\hat{\mu}}\hat{\psi} \ ,
\end{split}\end{align}
where $\hat{F}_{\hat{\mu}\hat{\nu}}=\partial_{\hat\mu}{\hat V}_{\hat\nu} -\partial_{\hat\nu}{\hat V}_{\hat\mu}\,$.

In the following, we will split the 4D coordinates as $x^{\hat \mu}=(x^\mu,x^3)$, where $x^3$ denotes the compactified
circle coordinate. Since all fields are periodic in $x^3$, we can write them as a Fourier series. For example:
\begin{align}\label{KKexpand}
 \hat{V}_{\hat{\mu}}(x^{\hat\mu})=\sum_n V_{\hat\mu,n}(x^\mu) e^{inmx^3}
 \,,\hskip 2truecm n \in \mathbb{Z}\ ,
\end{align}
where $m\ne 0$ has mass dimensions and corresponds to the inverse circle radius. The Fourier coefficients $V_{\hat\mu,n}(x^\mu)$
correspond to three-dimensional (un-hatted) fields. We first consider the bosonic fields. The reality condition on the 4D vector
and scalar implies that only the 3D ($n=0$) zero modes are real. All other modes are complex but only the positive ($n \ge 1$)
modes are independent, since
\begin{align}
V_{\hat\mu, -n} = V_{\hat\mu, n}^\star\ ,\hskip 1truecm F_{-n} =F_n^\star\ ,\hskip 1truecm n\ne 0\ .
\end{align}
In the following we will be mainly interested in the $n=1$ modes whose real and imaginary parts we indicate by
\begin{align}
V_\mu^{(1)} &\equiv \frac12\big(V_{\mu, 1} + V_{\mu,1}^\star\big)\ ,\hskip 1.1truecm
 V_\mu^{(2)} \equiv \frac{1}{2i}\big(V_{\mu, 1} - V_{\mu, 1}^\star\big)\ ,\nonumber\\[.1truecm]
\phi^{(1)} &\equiv \frac{1}{2}\big(V_{3, 1} + V_{3,1}^\star\big)\ ,\hskip 1.3truecm
 \phi^{(2)} \equiv \frac{1}{2i}\big(V_{3, 1} - V_{3, 1}^\star\big)\ ,\\[.1truecm]
F^{(1)} &\equiv \frac{1}{2}\big(F_{ 1} + F_{ 1}^\star\big)\ ,\hskip1.6truecm
 F^{(2)} \equiv \frac{1}{2i}\big(F_{ 1} - F_{1}^\star\big)\ .\nonumber
\end{align}

Similarly, the Majorana condition of the 4D spinor $\hat\psi$ implies that the $n=0$ mode is Majorana but that the independent
positive ($n\ge1$) modes are Dirac. This is equivalent to two (4-component, 3D reducible) Majorana spinors which we indicate by
\begin{align}
 \psi^{(1)}=\frac{1}{2}\big(\psi_1+B^{-1}\psi_1^\star\big) \ ,\qquad\qquad
 \psi^{(2)}=\frac{1}{2i}\big(\psi_1-B^{-1}\psi_1^\star\big)\ .
\end{align}
Here $B$ is the $4\times 4$ matrix $B= i C\Gamma_0$, where $C$ is the $4\times 4$ charge conjugation matrix.

Substituting the harmonic expansion \eqref{KKexpand} of the fields and a similar expansion of the gauge parameter
$\hat\Lambda$  into the transformation rules \eqref{v4rules}, we find the following transformation rules for the first
($n=1$) KK modes:
\begin{align}\begin{split}\label{KKv4rules}
 \delta \phi^{(1)}  &= -\bar{\epsilon}\Ga_3 \psi^{(1)}-m\Lambda^{(2)}-m\xi\phi^{(2)}\ ,\\[.2truecm]
     \delta \phi^{(2)} &= -\bar{\epsilon}\Ga_3 \psi^{(2)} +m\Lambda^{(1)} +m\xi\phi^{(1)}\ , \\[.2truecm]
 \delta V_\mu^{(1)} &= -\bar{\epsilon}\Ga_\mu\psi^{(1)} +\partial_\mu\Lambda^{(1)}-m\xi V_\mu^{(2)}\ ,\\[.2truecm]
 \delta V_\mu^{(2)} &= -\bar{\epsilon}\Ga_\mu\psi^{(2)} +\partial_\mu\Lambda^{(2)}+m\xi V_\mu^{(1)}\ ,\\[.2truecm]
 \delta F^{(1)} &= i\bar{\epsilon}\Ga_5\Ga^\mu\partial_\mu \psi^{(1)}
      -im\bar{\epsilon}\Ga_5\Ga_3\psi^{(2)}-m\xi F^{(2)}\ , \\[.2truecm]
 \delta F^{(2)} &= i\bar{\epsilon}\Ga_5\Ga^\mu\partial_\mu\psi^{(2)}
      +im\bar{\epsilon}\Ga_5\Ga_3\psi^{(1)} +m\xi F^{(1)}\ , \\[.2truecm]
 \delta \psi^{(1)} &= \frac18\Ga^{\mu\nu} F_{\mu\nu}^{(1)}\epsilon +\frac14\Ga^\mu\Ga_3\partial_\mu\phi^{(1)} \epsilon
          +\frac{i}{4}\Ga_5 F^{(1)}\epsilon+\frac{m}{4}\Ga^\mu\Ga_3 V_\mu^{(2)}\epsilon -m\xi \psi^{(2)}\ , \\[.2truecm]
 \delta \psi^{(2)}&= \frac18\Ga^{\mu\nu} F_{\mu\nu}^{(2)}\epsilon +\frac14\Ga^\mu\Ga_3\partial_\mu\phi^{(2)}\epsilon
          +\frac{i}{4}\Ga_5 F^{(2)}\epsilon-\frac{m}{4}\Ga^\mu\Ga_3 V_\mu^{(1)}\epsilon + m\xi \psi^{(1)}\ ,
\end{split}\end{align}
where we have defined
\begin{align}
\Lambda^{(1)} = \frac{1}{2}\big(\Lambda_1+\Lambda_1^\star\big)\ ,\hskip 1truecm
\Lambda^{(2)} = \frac{1}{2i}\big(\Lambda_1-\Lambda_1^\star\big)\ .
\end{align}
Apart from global supersymmetry transformations with parameter $\epsilon$ and gauge transformations with parameters
$\Lambda^{(1)}$, $\Lambda^{(2)}$, the transformations \eqref{KKv4rules} also contain a global $\mathrm{SO}(2)$ transformation
with parameter $\xi$, that rotates the real and imaginary parts of the 3D fields. This $\mathrm{SO}(2)$ transformation corresponds
to a central charge transformation and is a remnant of the translation in the compact circle direction.\footnote{This is a
conventional central charge transformation. Three-dimensional supergravity also allows for non-central charges from
extensions by non-central R-symmetry generators \cite{Nahm:1977tg}, recently discussed in \cite{Bergshoeff:2008ta}.}

In order to write the 3D 4-component Majorana spinors in terms of two irreducible 2-component Majorana spinors it is convenient to
choose the following representation of the $\Gamma$-matrices in terms of $2\times2$ block matrices:
\begin{align}\label{Gammas}
 \Ga_\mu=\bigg(\begin{matrix}\ga_\mu & 0\\0&-\ga_\mu\end{matrix}\bigg) \ ,\quad
 \Ga_3=\bigg(\begin{matrix}0 & \mathbf{1}\\\mathbf{1}&0\end{matrix}\bigg) \ ,\quad
 \Ga_5=\bigg(\begin{matrix}0&-i\\i&0\end{matrix}\bigg) \ .
\end{align}
The 3D $2\times 2$ matrices  $\ga_\mu$ satisfy the standard relations $\{\ga_\mu,\ga_\nu\}=2\eta_{\mu\nu}$ and can be chosen explicitly in terms of the Pauli matrices by
\begin{align}
\ga_\mu=(i\sigma_1,\sigma_2,\sigma_3) \,.
\end{align}
In this representation the 4D
charge conjugation matrix $C$ is given by
\begin{align}
 C=\bigg(\begin{matrix}\varepsilon&0\\0&-\varepsilon\end{matrix}\bigg)  \ ,
\end{align}
where
\begin{align}
 \varepsilon=\bigg(\begin{matrix}0&1\\-1&0\end{matrix}\bigg)
\end{align}
is the 3D charge conjugation matrix.

Using the above representation the 4-component Majorana spinors decompose into two 3D irreducible Majorana spinors as follows:
\begin{align}
\psi^{(1)}=\bigg(\begin{matrix} \chi_1\\ \chi_2\end{matrix}\bigg)\ , \qquad
 \psi^{(2)}=\bigg(\begin{matrix}\psi_1\\\psi_2\end{matrix}\bigg) \ , \qquad
 \epsilon=\bigg(\begin{matrix}\epsilon_1\\\epsilon_2\end{matrix}\bigg) \ .
\end{align}
In terms of these 2-component spinors the transformation rules \eqref{KKv4rules} read
\begin{align}\begin{split}\label{toomanymassiverules}
 \delta \phi^{(1)} &= -\bar{\epsilon}_1\chi_2+\bar{\epsilon}_2\chi_1 -m\Lambda^{(2)}-m\xi\phi^{(2)}\ ,\\[.2truecm]
 \delta \phi^{(2)} &= -\bar{\epsilon}_1\psi_2+\bar{\epsilon}_2\psi_1 +m\Lambda^{(1)}+m\xi\phi^{(1)}\ , \\[.2truecm]
 \delta  V_\mu^{(1)} &= -\bar{\epsilon}_1\ga_\mu \chi_1-\bar{\epsilon}_2\ga_\mu \chi_2 +\partial_\mu \Lambda^{(1)}
 -m\xi V_\mu^{(2)}\ ,\\[.2truecm]
 \delta  V_\mu^{(2)} &= -\bar{\epsilon}_1\ga_\mu\psi_1-\bar{\epsilon}_2\ga_\mu\psi_2 +\partial_\mu
 \Lambda^{(2)} +m\xi V_\mu^{(1)}\ , \\[.2truecm]
 \delta F^{(1)} &= -\bar{\epsilon}_1\ga^\mu\partial_\mu \chi_2+\bar{\epsilon}_2\ga^\mu\partial_\mu \chi_1
                 -m(\bar{\epsilon}_1\psi_1+\bar{\epsilon}_2\psi_2)
                 -m\xi F^{(2)} \ ,\\[.2truecm]
 \delta  F^{(2)} &= -\bar{\epsilon}_1\ga^\mu\partial_\mu\psi_2+\bar{\epsilon}_2\ga^\mu\partial_\mu\psi_1
                 +m(\bar{\epsilon}_1\chi_1+\bar{\epsilon}_2\chi_2)
                 +m\xi F^{(1)}\ , \\[.2truecm]
 \delta \chi_1 &= \frac18\ga^{\mu\nu} F_{\mu\nu}^{(1)}\epsilon_1+\frac14\big(
                 \ga^\mu\partial_\mu\phi^{(1)}+ F^{(1)} +m\ga^\mu V_\mu^{(2)}\big)\epsilon_2 -m\xi\psi_1\ ,\\[.2truecm]
 \delta \chi_2 &= \frac18\ga^{\mu\nu} F_{\mu\nu}^{(1)}\epsilon_2-\frac14\big(
                 \ga^\mu\partial_\mu \phi^{(1)}+ F^{(1)} +m\ga^\mu V_\mu^{(2)}\big)\epsilon_1 -m\xi\psi_2\ ,\\[.2truecm]
 \delta \psi_1 &= \frac18\ga^{\mu\nu}F_{\mu\nu}^{(2)}\epsilon_1+\frac14\big(
                 \ga^\mu\partial_\mu \phi^{(2)}+  F^{(2)} -m\ga^\mu  V_\mu^{(1)}\big)\epsilon_2  +m\xi \chi_1\ ,\\[.2truecm]
 \delta \psi_2 &= \frac18\ga^{\mu\nu} F_{\mu\nu}^{(2)}\epsilon_2-\frac14\big(
                 \ga^\mu\partial_\mu \phi^{(2)}+  F^{(2)} -m\ga^\mu  V_\mu^{(1)}\big)\epsilon_1+m\xi \chi_2 \ .
\end{split}\end{align}

If we take $m\to0$ in the above multiplet we obtain two decoupled multiplets, ($\phi^{(1)}$, $V_\mu^{(1)}$, $F^{(1)}$,
$\chi_1$, $\chi_2$) and $(\phi^{(2)},V_\mu^{(2)},F^{(2)},\psi_1,\psi_2)$. Either one of them constitutes a massless
$\mathcal{N}=2$ vector multiplet. This massless limit has to be distinguished from the massless limits discussed in
subsections \ref{sec2.3} and \ref{sec3.2}, which refer to limits taken after truncating to $\mathcal{N}=1$ supersymmetry.

\subsection{Truncation}

In the process of KK reduction, the number of supercharges stays the same. The  3D multiplet \eqref{toomanymassiverules}
we found in the previous subsection thus exhibits four supercharges and hence corresponds to an $\mathcal{N}=2$ multiplet,
containing two vectors and a central charge transformation. One can, however, truncate it to an $\mathcal{N}=1$ multiplet,
not subjected to a central charge transformation and containing only one vector. This truncated multiplet will be the
starting point to obtain an $\mathcal{N}=1$ supersymmetric version of the Proca theory. The $\mathcal{N}=1$ truncation is
given by:
\begin{align}
\phi^{(2)} = V_\mu^{(1)} = F^{(2)} = \chi_2 = \psi_1 = 0\ ,
\end{align}
provided that at the same time we truncate the following symmetries:
\begin{align}
\epsilon_1 = \Lambda^{(1)} = \xi = 0\ .
\end{align}
Substituting this truncation into the transformation rules \eqref{toomanymassiverules}, we find the following $\mathcal{N}=1$
massive vector supermultiplet: \footnote{Note that the field content given in \eqref{noeps1} is that of massless
$\mathcal{N}=2$. In the massive case, however, the scalar field $\phi$ will disappear after gauge-fixing the
St\"uckelberg symmetry.}
\begin{align}\begin{split}\label{noeps1}
 \delta \phi^{(1)} &= \bar{\epsilon}_2\chi_1 -m\Lambda^{(2)}\ ,\\[.1truecm]
 \delta V_\mu^{(2)} &= -\bar{\epsilon}_2\ga_\mu\psi_2 +\partial_\mu\Lambda^{(2)}\ , \\[.1truecm]
 \delta \psi_2 &= \frac18\ga^{\mu\nu}F_{\mu\nu}^{(2)}\epsilon_2 \ , \\[.1truecm]
 \delta \chi_1 &= \frac14\big(\ga^\mu\partial_\mu\phi^{(1)}+  F^{(1)} +m\ga^\mu  V_\mu^{(2)}\big)\epsilon_2 \ ,\\[.1truecm]
 \delta F^{(1)} &= \bar{\epsilon}_2\ga^\mu\partial_\mu \chi_1 -m\bar{\epsilon}_2\psi_2 \ .
\end{split}\end{align}
Redefining $\epsilon_2\to \epsilon\,, \Lambda^{(2)}\to \Lambda$ and
\begin{align}
\phi^{(1)}\to 4\phi\,,\quad  V_\mu^{(2)}\to V_\mu \ ,\quad F^{(1)}\to-F\,,\quad
\psi_2\to\psi\,,\quad \chi_1\to\chi \quad{\rm and}\quad m\to4m\ ,
\end{align}
we obtain
\begin{align}\begin{split}\label{vectorwithoutgaugefix}
 \delta\phi &= \frac14\bar{\epsilon}\chi -m\Lambda\ ,\\[.1truecm]
 \delta V_\mu &= -\bar{\epsilon}\ga_\mu\psi +\partial_\mu\Lambda\ ,\\[.1truecm]
 \delta \psi &= \frac18\ga^{\mu\nu}F_{\mu\nu}\epsilon \ ,\\[.1truecm]
 \delta \chi &= \ga^\mu D_\mu\phi\,\epsilon-\frac14F\epsilon \ ,\\[.1truecm]
 \delta F &= -\bar{\epsilon}\ga^\mu\partial_\mu\chi+4m\bar{\epsilon}\psi \ ,
\end{split}\end{align}
where the covariant derivative $D_\mu$ is defined as
\begin{align}
 D_\mu\phi=\partial_\mu\phi+m V_\mu \ .
\end{align}
The transformation rules \eqref{vectorwithoutgaugefix} leave the following action invariant:
\begin{align}\label{actionPbefore}
I_1=\int d^3x\left(-\frac{1}{4}F_{\mu\nu}\,F^{\mu\nu}-\frac12\,D_\mu\phi\,D^\mu\phi
  -2\bar\psi\,\partial \hskip -.22truecm /\,\psi
  -\frac{1}{8}\bar\chi\,\partial \hskip -.22truecm /\,\chi+\,m\,\bar\psi\chi+\frac{1}{32}F^2\right)\ .
\end{align}

The gauge transformation with parameter $\Lambda$ is a St\"uckelberg symmetry, that can be fixed by imposing the gauge condition
\begin{align}\label{gaugefix}
\phi = \text{const}\ .
\end{align}
Taking the resulting compensating gauge transformation
\begin{align}
\Lambda = \frac{1}{4m}\,\bar\epsilon\chi
\end{align}
into account, we obtain the final form of the supersymmetry transformation rules of the $\mathcal{N}=1$ supersymmetric Proca theory:
\begin{align}\begin{split}\label{massivevector}
 \delta V_\mu&=-\bar{\epsilon}\ga_\mu\psi+\frac{1}{4m}\bar{\epsilon}\partial_\mu\chi \ ,\\[.1truecm]
 \delta \psi &= \frac18\ga^{\mu\nu}F_{\mu\nu}\epsilon \ ,\\[.1truecm]
 \delta\chi &= m\ga^\mu\epsilon V_\mu-\frac14F\epsilon \ ,\\[.1truecm]
 \delta F &= -\bar{\epsilon}\ga^\mu\partial_\mu\chi +4m\bar{\epsilon}\psi \ .
\end{split}\end{align}
The supersymmetric Proca action is then given by
\begin{align}\label{SProcaAction}
I_{\mathrm{Proca}}=\int d^3x\left(- \frac{1}{4}F_{\mu\nu}\,F^{\mu\nu}
-\frac{1}{2}\,m^2V_\mu\,V^\mu -2\bar\psi\,\partial \hskip -.22truecm
/\,\psi -\frac{1}{8}\bar\chi\,\partial \hskip -.22truecm
/\,\chi+\,m\,\bar\psi\chi+\frac{1}{32}F^2\right)\ .
\end{align}
The supersymmetric Proca theory describes 2+2  on-shell and 4+4 off-shell degrees of freedom. 

This finishes our description of how to obtain the 3D off-shell massive $\mathcal{N}=1$ vector multiplet from a KK reduction and
subsequent truncation onto the first massive KK sector of the 4D off-shell massless $\mathcal{N}=1$ vector multiplet.

\subsection{Massless limit}\label{sec2.3}

We end this section with some comments on the massless limit ($m \rightarrow 0$). Taking the massless limit in
\eqref{vectorwithoutgaugefix}, we see that the Proca multiplet splits into a massless vector multiplet and a massless
scalar multiplet. Note that a massless vector multiplet can be coupled to a current supermultiplet. This  is a feature that we
would like to incorporate, in view of the upcoming spin-2 discussion. We will do so by coupling the above supersymmetric
Proca system to a conjugate multiplet $(J_\mu, \mathcal{J}_\psi, \mathcal{J}_\chi, J_F)$, where $J_\mu$ is a vector,
$\mathcal{J}_\psi$ and $\mathcal{J}_\chi$ are spinors and $J_F$ is a scalar. Our starting point is then the action
\begin{align} \label{currentaction}
I = I_{\mathrm{Proca}} + I_{\mathrm{int}}\ ,
\end{align}
where the interaction part $I_{\mathrm{int}}$ describes the coupling between the Proca multiplet and the conjugate multiplet:
\begin{align}
I_{\mathrm{int}} = V^\mu J_\mu + \bar{\psi} \mathcal{J}_\psi + \bar{\chi} \mathcal{J}_\chi + F J_F \ .
\end{align}
Requiring that $I_{\mathrm{int}}$ is separately invariant under supersymmetry, determines the transformation rules of the
conjugate multiplet:
\begin{align}\begin{split} \label{transconj}
\delta J_\mu & = \frac14\, \bar{\epsilon}\, \gamma_{\mu \nu} \partial^\nu \!\mathcal{J}_\psi
                 + m \bar{\epsilon}\, \gamma_\mu \mathcal{J}_\chi \ , \\[.1truecm]
\delta \!\mathcal{J}_\psi & = - \gamma^\mu \epsilon\, J_\mu - 4 m \epsilon\, J_F \ ,\\[.1truecm]
\delta \!\mathcal{J}_\chi & = \frac{1}{4 m}\, \epsilon\, \partial^\mu J_\mu + \gamma^\mu \epsilon\, \partial_\mu J_F \ , \\[.1truecm]
\delta J_F & = \frac14\, \bar{\epsilon} \mathcal{J}_\chi \ .
\end{split}\end{align}
Taking the massless limit in the action \eqref{currentaction} and transformation rules \eqref{massivevector}, \eqref{transconj}
is non-trivial, due to the factors of $1/m$ that appear in the transformation rules. In order to be able to take the limit in a
well-defined fashion, we will work in the formulation where the St\"uckelberg symmetry is not yet fixed. Note that this
formulation can be easily retrieved from the gauge fixed version, by making the following redefinition in the action
\eqref{SProcaAction} and transformation rules \eqref{massivevector}:
\begin{align}\label{redef}
V_\mu= \tilde V_\mu+\frac1m\partial_\mu\phi\ .
\end{align}
Applying this redefinition to \eqref{SProcaAction} and \eqref{massivevector} indeed brings one back to the action
\eqref{actionPbefore} and to the transformation rules \eqref{vectorwithoutgaugefix}, whose massless limit is well-defined.
The massless limit of the interaction part $I_\mathrm{int}$ (after performing the above substitution) and of the transformation
rules \eqref{transconj} of the conjugate multiplet, is however not well-defined. In order to remedy this, we will impose
the constraint that $J_\mu$ corresponds to a conserved current, i.e.~that
\begin{equation}
\partial^\mu J_\mu = 0 \ .
\end{equation}
In order to preserve supersymmetry, we will also take
$\mathcal{J}_\chi = 0$ and $J_F = 0$.\footnote{Strictly speaking,
preservation of the constraint $\partial^\mu J_\mu = 0$ under
supersymmetry leads to the constraint $\slashed \partial
\!\mathcal{J}_\chi = 0$ and preservation of this new constraint
leads to the constraint $\Box J_F = 0$. We are however interested in
the massless limit, in which the conserved currents  $(J_\mu,
\mathcal{J}_\psi)$ and the fields $(\mathcal{J}_\chi,J_F)$ form two
separate multiplets, that couple to a massless vector and scalar
multiplet respectively. Since we are mostly interested in the
coupling of the supercurrent multiplet $(J_\mu, \mathcal{J}_\psi)$
to the vector multiplet, we will simply set the fields
$(\mathcal{J}_\chi,J_F)$ equal to zero.} The conjugate multiplet
then reduces to a spin-1 current supermultiplet.

The massless limit is now everywhere well-defined. The transformation rules \eqref{vectorwithoutgaugefix} reduce to the
transformation rules of a massless vector $(\tilde{V}_\mu,\psi)$ and
scalar $(\phi,\chi,F)$ multiplet, see eqs.~\eqref{masslessV} and \eqref{masslessS},
respectively. Performing the above outlined procedure and taking the limit $m\rightarrow 0$ leads to the following action
\begin{align}
I  = \int d^3x\Big[\left(- \frac14\tilde F_{\mu\nu}\,\tilde F^{\mu\nu}
  -2\bar\psi\,\partial \hskip -.22truecm /\,\psi + \tilde{V}^\mu J_\mu + \bar{\psi}\mathcal{J}_\psi \right) \nonumber \\
  -\frac{1}{2}\left(\partial_\mu\phi\,\partial^\mu\phi
  +\frac{1}{4}\bar\chi\,\partial \hskip -.22truecm /\,\chi  -\frac{1}{16}F^2\right) \Big]\ ,
\end{align}
which is the sum of the supersymmetric massless vector and scalar multiplet actions, see eqs.~\eqref{masslessactionV} and
\eqref{masslessactionS}, respectively. The vector multiplet action is coupled to a spin-1 current multiplet. Note that
there is no coupling left between the current multiplet and the scalar multiplet. This will be different in the spin-2 case,
as we will see later.

%%%%%%%%%%%%%%
\section{Supersymmetric Fierz--Pauli}\label{sec:sec3}

In this section we extend the discussion of the previous section to the spin-2 case, skipping some of
the details  we explained in the spin-1 case. We use the same notation.

\subsection{Kaluza--Klein reduction and truncation}

Our starting point is the off-shell 4D $\mathcal{N}=1$ massless
spin-2 multiplet which consists of a symmetric tensor $\hat h_{\hat
\mu \hat\nu}\,$, a gravitino $\hat\psi_{\hat\mu}\,$, an auxiliary
vector $\hat A_{\hat\mu}$ and two auxiliary scalars $\hat  M$ and
$\hat N$. This corresponds to the linearized version of the `old
minimal supergravity' multiplet. The supersymmetry rules, with
constant spinor parameter $\epsilon$, and gauge transformations of
these fields, with local vector parameter ${\hat \Lambda}_{\hat
\mu}$ and local spinor parameter $\hat \eta$, are given by
\cite{Stelle:1978ye,Ferrara:1978em}:
\begin{align}
\delta\hat h_{\hat \mu\hat\nu} &={\bar\epsilon}\,\Gamma_{(\hat \mu}\,\hat\psi_{\hat\nu)}
          +\partial_{(\hat\mu}\,\hat\Lambda_{\hat\nu)}\ ,\nonumber\\[.1truecm]
\delta \hat\psi_{\hat\mu} &= -\frac14\Gamma^{\hat\rho\hat\lambda}\,\partial_{\hat\rho} \hat h_{\hat \lambda \hat\mu}\epsilon
    -\frac{1}{12}\Gamma_{\hat \mu}(\hat M+i\,\Gamma_5\hat N)\,\epsilon
    +\frac{1}{4}i\, \hat A_{\hat \mu}\Gamma_5\epsilon
    -\frac{1}{12}i\,\Gamma_{\hat\mu}\Gamma^{\hat\rho}\hat A_{\hat \rho}\Gamma_5\epsilon
    +\partial_{\hat\mu} \hat\eta\ ,\nonumber\\[.1truecm]
\delta \hat M &= -{\bar\epsilon} \,\Gamma^{\hat \rho\hat\lambda}\,\partial_{\hat\rho}\hat \psi_{\hat\lambda}
    \ ,\label{4DmasslessS=2}\\[.1truecm]
\delta \hat N &= -i\,{\bar\epsilon}\,\Gamma_5\,\Gamma^{\hat\rho\hat\lambda}\,\partial_{\hat\rho}\hat \psi_{\hat\lambda}\ ,\nonumber\\[.1truecm]
\delta \hat A_{\hat\mu} &= \frac32i \,\bar\epsilon\,\Gamma_5\,\Gamma_{\hat \mu}^{\;\;\hat\rho\hat\lambda}\,
\partial_{\hat\rho}\,\hat \psi_{\hat\lambda}
    -i\,{\bar\epsilon}\,\Gamma_5\,\Gamma_{\hat\mu}\Gamma^{\hat \rho\hat\lambda}\,
\partial_{\hat\rho}\hat \psi_{\hat\lambda}\ .\nonumber
\end{align}

Like in the spin-1 case we first perform a harmonic expansion of all fields and local parameters
and substitute these into the transformation rules \eqref{4DmasslessS=2}. Projecting onto the lowest KK massive sector
we then obtain all the transformation rules  of the real and imaginary parts of the $n=1$ modes, like in eq.~\eqref{KKv4rules}
for the spin-1 case. We indicate the real and imaginary parts of the bosonic modes by:
\begin{align}\begin{split}
h_{\mu\nu}^{(1)}&\equiv\frac{1}{2}\left(h_{\mu\nu,1}+h_{\mu\nu,1}^\star\right)\ ,\hskip 1truecm
    h_{\mu\nu}^{(2)}\equiv\frac{1}{2i}\left(h_{\mu\nu,1}-h_{\mu\nu,1}^\star\right)\ ,\\[.1truecm]
V_{\mu}^{(1)}&\equiv\frac{1}{2}\left(h_{\mu3,1}+h_{\mu3,1}^\star\right)\ , \hskip .95truecm
    V_{\mu}^{(2)}\equiv\frac{1}{2i}\left(h_{\mu3,1}-h_{\mu3,1}^\star\right)\ ,\\[.1truecm]
\phi^{(1)}&\equiv\frac{1}{2}\left(h_{33,1}+h_{33,1}^\star\right)\ , \hskip 1.1truecm
    \phi^{(2)}\equiv\frac{1}{2i}\left (h_{33,1}-h_{33,1}^\star\right)\ ,\\[.1truecm]
A_{\mu}^{(1)}&\equiv\frac{1}{2}\left(A_{\mu,1}+A_{\mu,1}^\star\right)\ , \hskip 1.15truecm
    A_{\mu}^{(2)}\equiv\frac{1}{2i}\left(A_{\mu,1}-A_{\mu,1}^\star\right)\ ,\\[.1truecm]
P^{(1)}&\equiv\frac{1}{2}\left(A_{3,1}+A_{3,1}^\star\right)\ , \hskip 1.25truecm
    P^{(2)}\equiv\frac{1}{2i}\left(A_{3,1}-A_{3,1}^\star\right)\ ,\\[.1truecm]
M^{(1)}&\equiv\frac{1}{2}\left(M_1+M_1^\star\right)\ , \hskip 1.45truecm
    M^{(2)}\equiv\frac{1}{2i}\left(M_1-M_1^\star\right)\ ,\\[.1truecm]
N^{(1)}&\equiv\frac{1}{2}\left(N_1+N_1^\star\right)\ , \hskip 1.65truecm
    N^{(2)}\equiv\frac{1}{2i}\left(N_1-N_1^\star\right)\ ,
\end{split}\end{align}
while  the fermionic modes decompose into two Majorana modes:
\begin{align}\begin{split}
\psi_{\mu}^{(1)}&\equiv\frac{1}{2}\left(\psi_{\mu,1}+B^{-1}\psi_{\mu,1}^\star\right)\ ,
\qquad
\psi_{\mu}^{(2)}\equiv\frac{1}{2i}\left(\psi_{\mu,1}-B^{-1}\psi_{\mu,1}^\star\right)\ ,\\[.1truecm]
\psi_{3}^{(1)}&\equiv\frac{1}{2}\left(\psi_{3,1}+B^{-1}\psi_{3,1}^\star\right)\ ,
\qquad \hskip .1truecm
\psi_{3}^{(2)}\equiv\frac{1}{2i}\left(\psi_{3,1}-B^{-1}\psi_{3,1}^\star\right)\ .
\end{split}\end{align}

We next use the representation \eqref{Gammas} of the $\Gamma$-matrices and decompose the 4-component spinors into two
2-component spinors as follows:
\begin{align}\begin{split}
\psi_\mu^{(1)}&= \left(\begin{array}{c} \psi_{\mu1} \\ \psi_{\mu2} \\ \end{array}\right)\,,
\qquad\quad
\psi_3^{(1)}= \left(\begin{array}{c} \chi_{1} \\ \chi_{2} \\ \end{array}\right)\,,\\[.1truecm]
\psi_\mu^{(2)}&= \left(\begin{array}{c} \chi_{\mu1} \\ \chi_{\mu2} \\ \end{array}\right)\,,
\qquad\quad
\psi_3^{(2)}= \left(\begin{array}{c} \psi_{1} \\ \psi_{2} \\ \end{array}\right)\,,\\[.1truecm]
\eta^{(1)}=\left(\begin{array}{c} \eta_{1}^{(1)} \\ \eta_{2}^{(1)} \\ \end{array}\right)\,&\,,
\qquad\quad
\eta^{(2)}= \left(\begin{array}{c} \eta_{1}^{(2)} \\ \eta_{2}^{(2)}\end{array}\right)\,, \hskip 1truecm
\epsilon =\left(\begin{array}{c} \epsilon_1 \\ \epsilon_2 \end{array}\right)\ .
\end{split}\end{align}
Furthermore, we perform the following consistent truncation of the fields \footnote{If we take the massless limit before
the mentioned truncation we find two copies of a $\mathcal{N}=2$ massless spin-2 multiplet plus two copies of a
$\mathcal{N}=2$ massless spin-1 multiplet, see also text after \eqref{toomanymassiverules}.}
\begin{align}
\phi^{(2)}= V_\mu^{(1)}= h_{\mu\nu}^{(2)}=M^{(2)}= N^{(1)}= P^{(2)}= A_{\mu}^{(1)}= \chi_2= \psi_1=\psi_{\mu1}=\chi_{\mu2}=0
\end{align}
and of the parameters
\begin{align}
\Lambda_\mu^{(2)}=\Lambda_3^{(1)}= \epsilon_1=\eta_1^{(1)} =\eta_2^{(2)}=\xi=0\ .
\end{align}
For simplicity, from now on we drop all numerical upper indices, e.g.~$\phi^{(1)}=\phi$, and all numerical lower indices,
e.g.~$\psi_{\mu 1}=\psi_\mu$ of the remaining non-zero fields (but not of the parameters). We find that the transformation rules of these
fields under supersymmetry, with constant 2-component spinor parameter $\epsilon$, and St\"uckelberg symmetries, with local scalar
and vector parameters $\Lambda_3\,, \Lambda_\mu\,,$ and 2-component spinor parameters
$\eta_1$ and $\eta_2$, are given by\,\footnote{The 4D analogue of this multiplet, in superfield language, can be
found in \cite{Buchbinder:2002gh}.}
\begin{align}\begin{split}\label{TransfRulesFP}
\delta h_{\mu\nu} &=\bar\epsilon\gamma_{(\mu}\psi_{\nu)}+\partial_{(\mu}\Lambda_{\nu)}\ ,\\[.1truecm]
\delta V_\mu &=\frac{1}{2}\bar\epsilon\gamma_\mu\psi -\frac{1}{2}\bar\epsilon\chi_{\mu} +\frac{1}{2}\partial_\mu\Lambda_3
   +\frac{1}{2}m\Lambda_\mu\ ,\\[.1truecm]
\delta\phi &=-\bar\epsilon\,\chi-m\Lambda_3\ ,\\[.1truecm]
\delta\psi_{\mu} &=-\frac{1}{4}\gamma^{\rho\lambda}\partial_\rho h_{\lambda\mu}\epsilon +\frac{1}{12}\gamma_\mu M\epsilon
   +\frac{1}{12}\gamma_\mu P\epsilon +\partial_\mu\eta_2\ ,\\[.1truecm]
\delta\psi &=-\frac{1}{4}\gamma^{\rho\lambda}\partial_\rho V_\lambda\epsilon -\frac{1}{12}N\epsilon
   -\frac{1}{12}\gamma^\rho A_\rho\epsilon +m\eta_2\ ,\\[.1truecm]
\delta \chi_{\mu} &=-\frac{1}{4}\gamma^\rho \partial_\rho V_\mu\epsilon +\frac{1}{4}m\gamma^\rho h_{\rho\mu}\epsilon
   -\frac{1}{12}\gamma_\mu N\epsilon +\frac{1}{4} A_\mu\epsilon -\frac{1}{12}\gamma_\mu\gamma^\rho A_\rho\epsilon
   +\partial_\mu\eta_1\ ,\\[.1truecm]
\delta\chi &=-\frac{1}{4}\gamma^\rho\partial_\rho\phi\epsilon -\frac{1}{12}M\epsilon+\frac{1}{6}P\epsilon
   -\frac{1}{4}m\gamma^\rho V_\rho\epsilon -m\eta_1\ ,\\[.1truecm]
\delta M &=-\bar\epsilon\gamma^\rho\partial_\rho\chi +\bar\epsilon\gamma^{\rho\lambda}\partial_\rho\psi_{\lambda}
   -m\bar\epsilon\gamma^\rho\chi_{\rho}\ ,\\[.1truecm]
\delta N &=-\bar\epsilon\gamma^\rho\partial_\rho\psi -\bar\epsilon\gamma^{\rho\lambda}\partial_\rho\chi_{\lambda}
   +m\bar\epsilon\gamma^\rho\psi_{\rho}\ ,\\[.1truecm]
\delta P &=\bar\epsilon\gamma^\rho\partial_\rho\chi +\frac{1}{2}\bar\epsilon\gamma^{\rho\lambda}\partial_\rho\psi_{\lambda}
   +m\bar\epsilon\gamma^\rho\chi_{\rho}\ ,\\[.1truecm]
\delta A_\mu &=\frac{3}{2}\bar\epsilon{\gamma_\mu}^{\rho\lambda}\partial_\rho\chi_{\lambda}
   -\bar\epsilon\gamma_\mu\gamma^{\rho\lambda}\partial_\rho\chi_{\lambda}
   +\frac{1}{2}\bar\epsilon{\gamma_\mu}^\rho\partial_\rho\psi -\bar\epsilon\partial_\mu\psi
   -\frac{1}{2}m\bar\epsilon{\gamma_\mu}^\rho\psi_{\rho} +m\bar\epsilon\psi_{\mu}\ .
\end{split}\end{align}
The action invariant under the transformations \eqref{TransfRulesFP} is given by
\begin{align}\label{maction}
  I_{m} =\int d^3x\Big\{& h^{\mu\nu} G^{\rm lin}_{\mu\nu}(h) -m^{2}\left( h^{\mu \nu }h_{\mu \nu }-h^{2}\right) \nonumber\\[.1truecm]
   +&  2h^{\mu\nu}\partial_\mu\partial_\nu\phi-2h\partial^\alpha\partial_\alpha\phi
      -F^{\mu\nu}F_{\mu\nu} +4m h^{\mu\nu}\partial_{(\mu}V_{\nu)}-4mh\partial^\mu V_\mu \nonumber\\[.2truecm]
   -&  4\bar\psi_\mu\gamma^{\mu\nu\rho}\partial_\nu\psi_\rho -4\bar\chi_\mu\gamma^{\mu\nu\rho}\partial_\nu\chi_\rho
       +8\bar\psi\gamma^{\mu\nu}\partial_\mu\chi_\nu +8\bar\psi_\mu\gamma^{\mu\nu}\partial_\nu\chi
       +8m\bar\psi_\mu\ga^{\mu\nu}\chi_\nu \nonumber\\[.2truecm]
   -&  \frac{2}{3}M^2 - \frac{2}{3}N^2+\frac{2}{3}P^2 + \frac{2}{3}A_\mu A^\mu \Big\}\ ,
\end{align}
where $h =\eta^{\mu\nu}h_{\mu\nu}$ and $G^{\rm lin}_{\mu\nu}(h)$ is the linearized Einstein tensor. We observe that the
action is non-diagonal in the bosonic fields
$(h_{\mu\nu}\,,V_\mu\,,\phi)$ and the fermionic fields $(\psi_\mu\,,\chi)$ and $(\chi_\mu\,,\psi)$.

Finally, we fix all St\"uckelberg symmetries by imposing the gauge conditions
\begin{align}\label{gauge conditions}
\phi ={\rm const}\ ,\qquad V_\mu=0\ ,\qquad\psi=0\ ,\qquad \chi=0\ .
\end{align}
Taking into account the  compensating gauge transformations
\begin{align}\begin{split}
\Lambda_3 &=0\ ,\\[.1truecm]
\Lambda_\mu &=\frac{1}{m}\bar\epsilon\chi_\mu\ ,\\[.1truecm]
\eta_1 &=-\frac{1}{12m}\left(M-2P\right)\epsilon\ ,\\[.1truecm]
\eta_2 &=\frac{1}{12m}\left(N+\gamma^\rho A_\rho\right)\epsilon\ ,
\end{split}\end{align}
we obtain the final form of the supersymmetry rules of the 3D $\mathcal N=1$ off-shell massive spin-2 multiplet:
\begin{align}\label{massivemultiplet}
\delta h_{\mu\nu} &=\bar\epsilon\gamma_{(\mu}\psi_{\nu)} +\frac{1}{m}\bar\epsilon\partial_{(\mu}\chi_{\nu)}\ ,\nonumber\\[.2truecm]
\delta\psi_{\mu} &=-\frac{1}{4}\gamma^{\rho\lambda}\partial_\rho h_{\lambda\mu}\epsilon
   +\frac{1}{12}\gamma_\mu (M+P)\epsilon +\frac{1}{12m}\partial_\mu(N+\gamma^\rho A_\rho)\epsilon\ ,\nonumber\\[.2truecm]
\delta \chi_{\mu} &= \frac{1}{4}m\gamma^\rho h_{\rho\mu}\epsilon +\frac{1}{4} A_\mu\epsilon
   -\frac{1}{12}\gamma_\mu (N+\gamma^\rho A_\rho)\epsilon -\frac{1}{12m}\partial_\mu(M-2P)\epsilon\ ,\nonumber\\[.2truecm]
\delta M &= \bar\epsilon\gamma^{\rho\lambda}\partial_\rho\psi_{\lambda}
   -m\bar\epsilon\gamma^\rho\chi_{\rho}\ ,\\[.2truecm]
\delta N &= -\bar\epsilon\gamma^{\rho\lambda}\partial_\rho\chi_\lambda +m\bar\epsilon\gamma^\rho\psi_\rho\ ,\nonumber\\[.2truecm]
\delta P &= \frac12\bar\epsilon\gamma^{\rho\lambda}\partial_\rho\psi_\lambda +m\bar\epsilon\gamma^\rho\chi_\rho\ ,\nonumber\\[.2truecm]
\delta A_\mu &=\frac{3}{2}\bar\epsilon{\gamma_\mu}^{\rho\lambda}\partial_\rho\chi_{\lambda}
   -\bar\epsilon\gamma_\mu\gamma^{\rho\lambda}\partial_\rho\chi_\lambda -\frac12m\bar\epsilon{\gamma_\mu}^\rho\psi_{\rho}
   +m\bar\epsilon\psi_{\mu}\ .\nonumber
\end{align}
These transformation rules leave the following action invariant:
\begin{align}\begin{split}\label{massiveaction}
I_{m\ne 0} =\int d^{3}x\,\Big\{ & h^{\mu\nu}G^{\text{lin}}_{\mu\nu}(h) -m^2\big(h^{\mu\nu}h_{\mu\nu}-h^2\big) \\[.2truecm]
 & -4\bar{\psi}_\mu \gamma^{\mu\nu\rho}\partial_\nu \psi_\rho  -4\bar{\chi}_\mu \gamma^{\mu\nu\rho}\partial_\nu \chi_\rho
   +8m\bar{\psi}_\mu \gamma^{\mu\nu}\chi_\nu \\[.2truecm]
 & -\frac{2}{3}M^2 -\frac{2}{3}N^2 +\frac{2}{3}P^2 +\frac{2}{3}A_\mu A^\mu\Big\}\ .
\end{split}\end{align}
This action describes 2+2 on-shell and 12+12 off-shell degrees of freedom.
The first line is the standard Fierz--Pauli action. The fermionic
off-diagonal mass term can easily be diagonalized by going to a basis in terms of the sum and difference of the two
vector-spinors.\footnote{The +3/2 and $-$3/2 helicity states are described by the sum and difference of the two
vector-spinors. See also appendix \ref{app:fermions}.}

The above action shows that the three scalars $M,\,N,\,P$ and the vector $A_\mu$ are auxiliary fields which are set to
zero by their equations of motion. We thus obtain the on-shell massive spin-2 multiplet with the following supersymmetry
transformations:
\begin{align}\begin{split}
\delta h_{\mu \nu} &= \bar{\epsilon} \gamma_{(\mu} \psi_{\nu)} + \frac{1}{m} \bar{\epsilon} \partial_{(\mu} \chi_{\nu)} \ , \\
\delta \psi_\mu &= -\frac14 \gamma^{\rho \sigma} \partial_\rho h_{\mu \sigma} \epsilon \ ,\\
\delta \chi_\mu &= \frac{m}{4} \gamma^\nu h_{\mu \nu} \epsilon \ .
\end{split}\end{align}

It is instructive to consider the closure of the supersymmetry algebra for the above supersymmetry rules  given the fact that,
unlike in the massless case, the symmetric tensor $h_{\mu\nu}$ does not transform under  the gauge transformations
$\delta h_{\mu\nu} = \partial_\mu\Lambda_\nu + \partial_\nu\Lambda_\mu$ and the only symmetries left to close the algebra are the global
translations. We find that the commutator of two supersymmetries on $h_{\mu\nu}$  indeed gives a translation,
\begin{align}\label{commutator}
[\delta_1, \delta_2] h_{\mu\nu}  = \xi^\rho \partial_\rho h_{\mu \nu} \ ,
\end{align}
with parameter
\begin{align}\label{translations}
\xi^\mu = \frac12 \bar{\epsilon}_2 \gamma^\mu \epsilon_1 \ .
\end{align}

To close the commutator on the two gravitini requires the use of the equations of motion for these fields. From the action
\eqref{massiveaction} we obtain the following equations:
\begin{align} \label{eomchi}
\gamma^{\mu \nu \rho} \partial_\nu \chi_\rho = m \gamma^{\mu \nu} \psi_\nu \ ,
\end{align}
and a similar equation for $\psi_\mu$. These equations of motion imply the standard spin-3/2 Fierz--Pauli equations
\begin{align}\begin{split}
\mathcal{R}_\mu^{(1)} &\equiv \partial \hskip -.24truecm / \,\chi_\mu + m \psi_\mu = 0 \ , \\[.2truecm]
\partial^\mu \chi_\mu &= 0 \ , \hskip 1.3truecm      \gamma^\mu \chi_\mu =  0 \ ,
\end{split}\end{align}
and similar equations for $\psi_\mu$. A useful alternative way of writing the equations of motion
\eqref{eomchi} is
\begin{align}
\mathcal{R}_{\mu \nu}^{(2)} \equiv \partial_{[\mu} \chi_{\nu]} + m \gamma_{[\mu} \psi_{\nu]} = 0 \ .
\end{align}
Using these two ways of writing the equations of motion as well as  the FP conditions  that follow from them we
find that the commutator on the two gravitini gives the same translations \eqref{translations} up to
equations of motion. More specifically, we find the  following commutators
\begin{align}\begin{split}
\left[\delta_1, \delta_2\right] \psi_\mu &= \xi^\nu\partial_\nu\psi_\mu
       -\frac{1}{4m}\xi^\alpha\partial_\mu\mathcal{R}_\alpha^{(1)}
       -\frac{1}{8m}\xi^\alpha\gamma_\alpha\partial_\mu(\gamma^{\rho\sigma} \partial_\rho \chi_\sigma) \\
 & \quad +\frac{1}{4m}\xi^\alpha\partial_\mu\partial_\alpha (\gamma^\sigma \chi_\sigma)
       -\frac{1}{8}\xi^\alpha\gamma_\mu \gamma_\alpha (\gamma^{\rho \sigma} \partial_\rho \psi_\sigma)\ , \\
\left[\delta_1, \delta_2\right] \chi_\mu &= \xi^\nu \partial_\nu \chi_\mu
        +\frac{1}{2} \xi^\nu\mathcal{R}_{\mu \nu}^{(2)} - \frac{1}{8} \xi^\rho \gamma_\rho \mathcal{R}_\mu^{(1)} \\
 & \quad -\frac18 \xi^\rho\gamma_\rho\partial_\mu(\gamma^\nu\chi_\nu) +\frac{m}{8}\xi^\rho\gamma_\rho\gamma_\mu(\gamma^\nu\psi_\nu) \ .
\end{split}\end{align}
Hence, the algebra closes on-shell.

\subsection{Massless limit}\label{sec3.2}

Finally, we discuss the massless limit $m\rightarrow 0$ of the supersymmetric FP theory. This is particularly interesting
in view of the fact that the massless limit of the ordinary spin-2 FP system, coupled to a conserved energy-momentum tensor
does not lead to linearized Einstein gravity. Instead, it leads to linearized Einstein gravity plus an extra force, mediated
by a scalar that couples to the trace of the energy-momentum tensor with gravitational strength. This phenomenon is known as the van Dam--Veltman--Zakharov
discontinuity. In the following, we will pay particular attention to this discontinuity in the supersymmetric case.

In order to discuss the massless limit, it turns out to be advantageous to trade the scalar fields $M$ and $P$ for scalars
$S$ and $F$, defined by
\begin{align}
 S=\frac{1}{6}\big(M+P\big)\ ,\qquad\qquad F=\frac43\big(M-2P\big)\ .
\end{align}
This field redefinition will make the multiplet structure of the resulting massless theory more manifest.
In order to discuss the vDVZ discontinuity, we will include a coupling to a conjugate multiplet
$(T_{\mu \nu}, \mathcal{J}_\mu^\psi, \mathcal{J}_\mu^\chi, T_S, T_N, T_F, T_\mu^A)$, as we did in the Proca case. Here
$T_{\mu \nu}$ is a symmetric two-tensor, $\mathcal{J}^\psi_\mu$, $\mathcal{J}^\chi_\mu$ are vector-spinors, $T_\mu^A$ is a
vector and $T_F$, $T_S$, $T_N$ are scalars. We will thus start from the action
\begin{align} \label{actintspin2}
I = I_{\rm FP} + I_{\mathrm{int}} \ ,
\end{align}
where $I_{\rm FP}$ is the supersymmetric FP action \eqref{massiveaction} and the interaction part $I_{\mathrm{int}}$ is given by
\begin{align} \label{intspin2}
I_{\mathrm{int}} = h_{\mu\nu}T^{\mu\nu} +\bar\psi_\mu\mathcal{J}^\mu_\psi +\bar\chi_\mu\mathcal{J}^\mu_\chi
     +S\,T_S +F\,T_F +N\,T_N + A_\mu\, T^\mu_A \ .
\end{align}
Requiring that $I_{\mathrm{int}}$ is separately invariant under supersymmetry determines the  transformation rules of the
conjugate multiplet:
\begin{align}\begin{split}\label{massivecurrents}
 \delta T_{\mu\nu}  &= \frac14 \bar\epsilon\,\ga_{\alpha (\mu}\partial^\alpha\!\mathcal{J}_{\nu)}^\psi
           +\frac{m}{4}\bar\epsilon\,\ga_{(\mu}\mathcal{J}_{\nu)}^\chi \ ,\\
 \delta \mathcal{J}_\mu^\psi &= \ga^\alpha\epsilon\,T_{\alpha\mu} +\frac14\ga_{\mu\alpha}\epsilon\,\partial^\alpha T_S
           +m\ga_\mu\epsilon\, T_N +\frac{m}{2}\ga_{\mu\alpha}\epsilon\, T^\alpha_A -m\epsilon\, T_\mu^A \ ,\\
 \delta \mathcal{J}_\mu^\chi &= \frac1m\epsilon\,\partial^\alpha T_{\mu\alpha}-\ga_{\mu\alpha}\epsilon\,\partial^\alpha T_N
           -4m\ga_\mu\epsilon\, T_F -\frac32\ga_{\mu\alpha\beta}\epsilon\,\partial^\alpha T^\beta_A
           +\ga_{\mu\alpha}\ga_\beta\epsilon\,\partial^\alpha T^\beta_A \ ,\\
 \delta T_S &= \frac12\bar\epsilon\,\ga^\mu\!\mathcal{J}_\mu^\psi \ ,\\
 \delta T_N &= \frac{1}{12m}\bar\epsilon\,\partial^\mu\!\mathcal{J}_\mu^\psi
           -\frac{1}{12}\bar\epsilon\,\ga^\mu\!\mathcal{J}_\mu^\chi \ ,\\
 \delta T_F &= -\frac{1}{16m}\bar\epsilon\,\partial^\mu\!\mathcal{J}_\mu^\chi \ ,\\
 \delta T_\mu^A &= -\frac14\bar\epsilon\mathcal{J}_\mu^\chi +\frac{1}{12}\bar\epsilon\,\ga_\mu\ga^\rho\!\mathcal{J}_\rho^\chi
           -\frac{1}{12m}\bar\epsilon\,\ga_\mu\partial^\rho\!\mathcal{J}_\rho^\psi \ .
\end{split}\end{align}
As in the Proca case, one should go back to a formulation that is
still invariant under the St\"uckelberg symmetries, in order to take
the massless limit in a well-defined way. This may be achieved by
making the following field redefinitions in the final transformation
rules \eqref{massivemultiplet} and action \eqref{massiveaction}
thereby re-introducing the  fields $(V_\mu\,,\phi'\,,\chi'\,,\psi)$
that were eliminated by the gauge-fixing conditions \eqref{gauge
conditions}:
\begin{align}
 h_{\mu\nu} &= \tilde h_{\mu\nu}-\frac1m\big(\partial_\mu V_\nu+\partial_\nu V_\mu\big)
              +\frac{1}{m^2}\partial_\mu\partial_\nu\phi'\ , \nonumber \\[.2truecm]
  \psi_\mu &=\tilde\psi_\mu-\frac1m\partial_\mu\psi \ ,\quad\quad \chi_\mu=\tilde\chi_\mu+\frac{1}{4m}\partial_\mu\chi' \ .
\end{align}
Applying this field redefinition in \eqref{massivemultiplet} then leads to transformation rules\footnote{These resulting
transformation rules are given by the transformation rules \eqref{TransfRulesFP}, provided one makes the following substitution:
$h_{\mu\nu}\rightarrow \tilde{h}_{\mu\nu}$, $\psi_\mu \rightarrow \tilde{\psi}_\mu$, $\chi_\mu\rightarrow \tilde{\chi}_\mu$,
$\phi \rightarrow -\phi'$ and $\chi \rightarrow \chi'/4$.}, whose massless limit is well-defined. In order to make the massless
limit of the interaction part $I_{\mathrm{int}}$ and of the transformation rules \eqref{massivecurrents} well-defined, we impose
that $T_{\mu \nu}$ and $\mathcal{J}^\psi_\mu$ are conserved
\begin{align}
\partial^\nu T_{\mu \nu} = 0 \ , \qquad \partial^\mu \!\mathcal{J}^\psi_\mu = 0 \ ,
\end{align}
and we put $\mathcal{J}^\chi_\mu$, $T_F$, $T_N$ and $T_\mu^A$ to
zero in order to preserve supersymmetry and to obtain an irreducible
multiplet in the massless limit. The conjugate multiplet
\eqref{massivecurrents} then reduces to a spin-2 supercurrent
multiplet $(T_{\mu \nu}, \mathcal{J}_\mu^\psi, T_S)$ that contains
the energy-momentum tensor $T_{\mu \nu}$ and supersymmetry current
$\mathcal{J}^\psi_\mu$.

As in the Proca case, the massless limit is now well-defined. Performing the above outlined steps on the action
\eqref{actintspin2} and taking the massless limit leads, however, to an action that is in off-diagonal form. This action can
be diagonalized by making the following field redefinitions:
\begin{align}\label{bosonicshift}
\tilde h_{\mu\nu} =h'_{\mu\nu} +\eta_{\mu\nu}\phi'\ ,\hskip .5truecm
\tilde\psi_\mu = \psi'_\mu+\frac14 \gamma_\mu \chi' \ ,\hskip .5truecm
S = S'-\frac18F \ ,\hskip .5truecm
\tilde\chi_\mu= \chi'_\mu-\gamma_\mu \psi \ .
\end{align}
The resulting action is given by
\begin{align}\begin{split}\label{diagmasslessaction}
I =\int d^{3}x\,\Big\{
 & h'^{\mu\nu}G^{\text{lin}}_{\mu\nu}(h') -4\bar\psi'_\mu \gamma^{\mu\nu\rho}\partial_\nu \psi'_\rho
   -8S'^2 + h'_{\mu \nu} T^{\mu \nu} + \bar{\psi}'^\mu \!\mathcal{J}^\psi_\mu + S' T_S \\[.2truecm]
 & -F^{\mu\nu}F_{\mu\nu}-\frac23N^2+\frac23A^\mu A_\mu -4\bar\chi'_\mu\ga^{\mu\nu\rho}\partial_\nu\chi'_\rho
   -8\bar\psi\ga^\mu\partial_\mu\psi  \\[.2truecm]
 & +2\big[-\partial_\mu\phi'\partial^\mu\phi'-\frac14\bar\chi'\ga^\mu\partial_\mu\chi'+\frac{1}{16}F^2\big] \\[.2truecm]
 & + \phi'\eta^{\mu\nu}T_{\mu\nu}-\frac14\bar\chi'\,\ga^\mu\!\mathcal{J}^\psi_\mu-\frac18F T_S \Big\}\ .
\end{split}\end{align}
This is an action for three massless multiplets : a spin two multiplet $(h'_{\mu\nu},\psi'_\mu,S')$, a mixed gravitino-vector
multiplet\footnote{An on-shell version of this multiplet was introduced in \cite{Nishino:2011zzd}.}
$(V_\mu,\chi'_\mu,\psi,N,A_\mu)$ and a scalar multiplet $(\phi',\chi',F)$. These multiplets and their transformation rules
are collected in appendix \ref{app:multiplets}.\footnote{The transformation rules of the different multiplets can also be
found by starting from the transformation rules of the massive FP multiplet and carefully following all redefinitions as
outlined in the main text, provided one performs compensating gauge transformations.} The spin-2 multiplet couples to the
supercurrent multiplet in the usual fashion. Unlike the Proca case however, the supercurrent multiplet does not only couple
to the spin-2 multiplet, but there is also a coupling to the scalar multiplet, given in the last line of
\eqref{diagmasslessaction}. Indeed, defining
\begin{align}
 T_\phi=\eta^{\mu\nu}T_{\mu\nu} \ , \qquad \mathcal{J}=-\frac14\ga^\mu\!\mathcal{J}_\mu^\psi\ ,\qquad T_F=-\frac18T_S \ ,
\end{align}
one finds that the fields $(T_\phi, \mathcal{J}, T_F)$ form a
conjugate scalar multiplet with transformation rules
\begin{align}\begin{split}\label{scalarcurrents}
 \delta T_\phi  &= -\bar\epsilon\,\ga^{\mu}\partial_\mu\mathcal{J} \ ,\\
 \delta\! \mathcal{J} &= -\frac14\epsilon\,T_\phi +\ga^\mu\epsilon\,\partial_\mu T_F \ ,\\
 \delta T_F &= \frac14\bar\epsilon\mathcal{J} \ ,
\end{split}\end{align}
such that the last line of \eqref{diagmasslessaction} is invariant under supersymmetry.

We have thus obtained a 3D supersymmetric version of the 4D vDVZ
discontinuity. The above discussion shows that the massless limit of
the supersymmetric FP theory coupled to a supercurrent multiplet,
leads to linearized $\mathcal{N}=1$ supergravity, plus an extra
scalar multiplet that couples to a multiplet that includes the trace
of the energy-momentum tensor and the gamma-trace of the
supercurrent.

\section{Linearized SNMG without Higher Derivatives}\label{sec:sec4}

Using the results of the previous section we will now construct linearized New Massive Supergravity without higher
derivatives but with auxiliary fields. Furthermore, we will show how, by eliminating the different ``non-trivial'' bosonic and
fermionic auxiliary fields, one re-obtains the higher-derivative kinetic terms for both the bosonic and fermionic fields.
We remind that by a ``non-trivial'' auxiliary field we mean an auxiliary field whose elimination leads to higher-derivative
terms in the action.

Consider first the bosonic case. The linearized version of lower-derivative (``lower'') NMG is
described by the following action \cite{Bergshoeff:2009hq}:
\begin{align}\label{linearizedNMG0}
I_{\text{NMG}}^{\text{lin }}(\text{lower}) =\int d^{3}x\,\Big\{ -h^{\mu\nu}G^{\text{lin}}_{\mu\nu}(h)
               + 2q^{\mu\nu}G^{\text{lin}}_{\mu\nu}(h) -m^2(q^{\mu\nu}q_{\mu\nu} - q^2)\Big\}\ ,
\end{align}
where $h_{\mu\nu}$ and $q_{\mu\nu}$ are two symmetric tensors and $q=\eta^{\mu\nu}q_{\mu\nu}$.
The above action can be diagonalized by making the redefinitions
\begin{align}
h_{\mu\nu} = A_{\mu\nu}+B_{\mu\nu}\ ,\qquad q_{\mu\nu} = B_{\mu\nu}\ ,\label{redef2}
\end{align}
after which we obtain
\begin{align}
I^{\text{lin}}_{\text{NMG}}\left[ A,B\right] =\int d^{3}x\,\Big\{- A^{\mu \nu}G^{\text{lin}}_{\mu \nu }\left( A\right)
            +B^{\mu \nu }G^{\text{lin}}_{\mu \nu }\left(B\right) -m^2\left( B^{\mu\nu}B_{\mu\nu}-B^2\right) \Big\} \ .
\end{align}
Using this diagonal basis it is clear that we can supersymmetrize the action in terms of a massless multiplet
$(A_{\mu\nu},\lambda_\mu,S)$ and a massive multiplet $(B_{\mu\nu},\psi_\mu,\chi_\mu,M,N,P,A_\mu) $.
Transforming this result back in terms of  $h_{\mu\nu}$ and $q_{\mu\nu}$ and making the redefinition
\begin{align}
\lambda_\mu = \rho_\mu-\psi_\mu
\end{align}
we find the following linearized lower-derivative supersymmetric NMG action
\begin{align}\label{linearizedNMG1}
I_{\text{SNMG}}^{\text{lin}}(\text{lower}) =\int d^{3}x\,&\Big\{-h^{\mu\nu}G^{\text{lin}}_{\mu\nu}(h)
     + 2q^{\mu\nu}G^{\text{lin}}_{\mu\nu}(h) -m^2(q^{\mu\nu}q_{\mu\nu} - q^2) +8S^2  \nonumber\\
  &  -\frac{2}{3}M^2 - \frac{2}{3}N^2+\frac{2}{3}P^2 + \frac{2}{3}A_\mu A^\mu \\[.2truecm]
  &  +4\bar{\rho}_\mu\gamma^{\mu\nu\rho}\partial_{\nu}\rho_\rho
     -8\bar{\psi}_\mu\gamma^{\mu\nu\rho}\partial_\nu\rho_\rho
     -4\bar{\chi}_\mu\gamma^{\mu\nu\rho}\partial_\nu\chi_\rho+8m\bar{\psi}_\mu\gamma^{\mu\nu}\chi_\nu \Big\} \ .\nonumber
\end{align}
This action describes 2+2 on-shell and 16+16 off-shell degrees of freedom.
It is invariant under the following transformation rules
\begin{align}\label{linearizedNMG2}
 \delta h_{\mu\nu} &=\bar{\epsilon}\gamma_{(\mu }\rho_{\nu)}\ , \nonumber \\[.1truecm]
 \delta \rho_\mu &= -\frac{1}{4}\gamma^{\rho\sigma}\big(\partial_\rho h_{\mu\sigma}\big)\epsilon
                     +\frac12 S\gamma_\mu\epsilon  +\frac{1}{12}\gamma_\mu (M+P)\epsilon \ ,\\[.1truecm]
 \delta S &= \frac14\bar \epsilon\gamma^{\mu\nu}\rho_{\mu\nu} -\frac14\bar\epsilon\gamma^{\mu\nu}\psi_{\mu\nu}\ ,\nonumber
\end{align}
where
\begin{align}
\rho_{\mu\nu} =\frac12\big(\partial_\mu\rho_\nu-\partial_\nu\rho_\mu\big)\ , \hskip 1truecm
   \psi_{\mu\nu} =\frac12\big(\partial_\mu\psi_\nu-\partial_\nu\psi_\mu\big)\ ,
\end{align}
plus the transformation rules for the massive multiplet $(q_{\mu\nu}\,,\psi_\mu\,,\chi_\mu\,,\,M,\,N,\,P,\,A_\mu)$
which can be found  in eq.~\eqref{massivemultiplet}, with $h_{\mu\nu}$ replaced by $q_{\mu\nu}$.
We have deleted $1/m$ terms in the transformation of $h_{\mu\nu}$ and $\rho_\mu$ since they take the form of a
gauge transformation. Note also that the auxiliary field $S$ transforms to the gamma trace of the
equation of motion for $\rho_\mu$.

The action \eqref{linearizedNMG1} contains the trivial auxiliary fields $(S,\,M,\,N,\,P,\,A_\mu)$ and the non-trivial auxiliary
fields $(q_{\mu\nu}\,, \psi_\mu\,, \chi_\mu)$. The elimination of the trivial auxiliary fields does not lead to anything new.
These fields  can simply be set equal to zero and disappear from the action.
Instead, as we will show now, the elimination of the non-trivial auxiliary fields
leads to higher-derivative terms in the action. To start with, the equation of motion for $q_{\mu\nu}$ can be used to
solve for $q_{\mu\nu}$ as follows:
\begin{align}\label{solqu}
q_{\mu \nu }=\frac{1}{m^{2}}G_{\mu\nu}^{\text{lin}}(h)-\frac{1}{2m^{2}} \eta_{\mu\nu}G^{\text{lin}}_{\text{tr}}(h)\ ,
\end{align}
where $G^{\text{lin}}_{\text{tr}}(h) = \eta^{\mu\nu}G_{\mu\nu}^{\text{lin}}(h)$. One of the vector-spinors, $\psi_\mu$, occurs
as a Lagrange multiplier. Its equation of motion enables one to solve for $\chi_\mu$:
\begin{align}\label{solchi}
\chi_\mu = -\frac{1}{2m}\gamma^{\rho\sigma}\gamma_\mu\,\rho_{\rho\sigma}\ .
\end{align}
The equation of motion of the other vector-spinor, $\chi_\mu$, can be used to solve for $\psi_\mu$ in terms of $\chi_\mu$:
\begin{align}\label{solpsi}
\psi_\mu = -\frac{1}{2m}\gamma^{\rho\sigma}\gamma_\mu\,\chi_{\rho\sigma}\ ,
\end{align}
and hence, via eq.~\eqref{solchi}, in terms of $\rho_\mu$. One can show that the solution of $\psi_\mu$ in terms of
(two derivatives of) $\rho_\mu$
is such that it solves the constraint
\begin{align}\label{constraintpsi}
\gamma^{\mu\nu}\psi_{\mu\nu}=0\ .
\end{align}
We now substitute the solutions \eqref{solqu} for $q_{\mu\nu}$ and \eqref{solchi} for $\chi_\mu$ back into the action
and make use of the identity
\begin{align}\label{identity1}
-4\bar\chi_\mu\gamma^{\mu\nu\rho}\partial_\nu\chi_\rho = \frac{8}{m^2}\bar\rho^{\mu\nu}\partial \hskip -.24truecm /\,\rho_{\mu\nu}
-\frac{2}{m^2}\bar\rho_{\mu\nu}\gamma^{\mu\nu}\partial \hskip -.24truecm /\,\gamma^{\sigma\rho}\rho_{\sigma\rho}\ ,
\end{align}
where we ignore a total derivative term. One thus obtains the following linearized higher-derivative (``higher'')
supersymmetric action of NMG \cite{Andringa:2009yc}:
\begin{align}\label{linearizedNMG3}
I^{\text{lin}}_{\text{SNMG}}(\text{higher}) &=\int d^{3}x\,\Big\{ -h^{\mu\nu}G^{\text{lin}}_{\mu\nu}(h)
     +4\bar{\rho}_\mu\gamma^{\mu\nu\rho}\partial_\nu \rho_\rho  +8S^2 \\
  &  +\frac{4}{m^2}\big(R^{\mu\nu}R_{\mu\nu}-\frac38 R^2\big)^{\text{lin}}
     +\frac{8}{m^2}\bar{\rho}_{ab}\partial \hskip -.24truecm /\,\rho_{ab}
     -\frac{2}{m^2} \bar{\rho}_{ab}\gamma^{ab}\partial \hskip -.24truecm / \,\gamma^{cd}\rho_{cd}\Big\}\nonumber \ .
\end{align}
The action \eqref{linearizedNMG3} is invariant under the supersymmetry rules
\begin{align}\label{linearizedNMG4}
\delta h_{\mu \nu } &=\bar{\epsilon}\gamma_{(\mu}\rho_{\nu)}\ ,  \notag \\[.1truecm]
\delta \rho_\mu &= -\frac{1}{4}\gamma^{\rho\sigma}\partial_\rho h_{\mu\sigma} \epsilon +\frac12 S\gamma_\mu\epsilon\ ,\\[.1truecm]
\delta S &= \frac14\bar\epsilon\gamma^{\mu\nu}\rho_{\mu\nu}\ ,\nonumber
\end{align}
where we made use of the constraint \eqref{constraintpsi} to simplify the transformation rule of $S$. Under supersymmetry
the auxiliary field $S$ transforms to the gamma-trace of the equation of motion for $\rho_\mu$, since  the higher-derivative
terms in this equation of motion are gamma-traceless and therefore drop out.

Alternatively, the higher-derivative kinetic terms for $\rho_\mu$ can be obtained by boosting up the derivatives in the
massive spin-3/2 FP equations in the same way as that has been done for the spin-2 FP equations in the
construction of New Massive Gravity  \cite{Bergshoeff:2009hq}, except for one subtlety, see appendix \ref{app:fermions}.

This finishes our construction of linearized SNMG. In the next section we will discuss to which extent this
result can be extended to the non-linear case.

\section{The non-linear case}\label{sec:sec5}

Supersymmetric NMG {\sl without} ``non-trivial'' auxiliary fields, i.e.~with higher derivatives, has already been
constructed some time ago \cite{Andringa:2009yc}. This action only contains the auxiliary field $S$ of the massless
multiplet. A characteristic feature is that there is no kinetic term for $S$ and in the bosonic terms $S$ occurs as
a torsion contribution to the spin-connection. However, due to its coupling to the fermions it cannot be eliminated
from the action. Thus, in the non-linear case we cannot anymore identify $S$ as a ``trivial'' auxiliary field.

We recall that, apart from the auxiliary field $S$, in the linearized analysis of section \ref{sec:sec3} and \ref{sec:sec4}
we distinguish between the trivial auxiliary fields $(M,\,N,\,P,\,A_\mu)$
and the non-trivial ones $(q_{\mu\nu}\,, \psi_\mu\,, \chi_\mu)$. Only the elimination of the latter ones leads to higher
derivatives in the Lagrangian. In the formulation of \cite{Andringa:2009yc} only the auxiliary field $S$ occurs.
One could now  search either for a formulation in which all other auxiliary fields occur or for an alternative
formulation in which only the non-trivial auxiliary fields $(q_{\mu\nu}\,, \psi_\mu\,, \chi_\mu)$ are present. In this
work we will not consider the inclusion of all auxiliary fields any further. It is not clear to us whether
such a formulation exists. This is based on the fact that our construction of the linearized massive multiplet makes
use of the existence of a consistent truncation to the first massive KK level. Such a truncation can
only be made consistently at the linearized level.

Before discussing the inclusion of the non-trivial auxiliary fields $(q_{\mu\nu}\,, \psi_\mu\,, \chi_\mu)$
it is instructive to first consider the linearized case and see how, starting from the (linearized) formulation of
\cite{Andringa:2009yc}  these three non-trivial auxiliary fields can be included and a formulation with lower
derivatives can be obtained. Our starting point is the higher-derivative action \eqref{linearizedNMG3} and corresponding
transformation rules \eqref{linearizedNMG4}. We first consider the bosonic part of the action \eqref{linearizedNMG3}, i.e.
\begin{align}\label{highNMG}
I_{\text{bos}}^{\text{lin}}(\text{higher}) =\int d^{3}x\,\Big\{-h^{\mu \nu }G_{\mu \nu }^{\text{lin}}\left( h\right) +8S^{2}
   +\frac{4}{m^2}\big(R^{\mu\nu}R_{\mu\nu}-\frac38 R^2\big)^{\text{lin}}\Big\}\ .
\end{align}
We already know from the construction of the bosonic theory that the derivatives can be lowered by introducing a symmetric auxiliary
field $q_{\mu\nu}$ and writing the equivalent bosonic action
\begin{align}\label{lowNMG}
I_{\text{bos}}^{\text{lin}}(\text{lower}) =\int d^{3}x\,\Big\{ -h^{\mu \nu }G_{\mu \nu }^{\text{lin}}\left( h\right) +8S^{2}
    +2q^{\mu \nu }G_{\mu \nu }^{\text{lin}}(h)-m^{2}\left( q^{\mu \nu}q_{\mu \nu }-q^{2}\right)  \Big\}\ .
\end{align}
The field equation of $q_{\mu\nu}$ is given by eq.~\eqref{solqu} and substituting this solution back into the
lower-derivative bosonic action \eqref{lowNMG} we re-obtain the higher-derivative bosonic action \eqref{highNMG}.

We next consider the fermionic part of the higher-derivative action \eqref{linearizedNMG3}, i.e.
\begin{align}\label{fermhigher}
I_{\text{ferm}}^{\text{lin}}(\text{higher}) =\int d^{3}x\,\Big\{4\bar{\rho}_{\mu }\gamma^{\mu\nu\rho}\partial_\nu \rho_\rho
       + \frac{8}{m^2}\bar{\rho}_{ab}\partial \hskip -.24truecm /\,\rho_{ab}
       -\frac{2}{m^2}\bar{\rho}_{ab}\gamma^{ab}\partial \hskip -.24truecm /\,\gamma^{cd}\rho_{cd} \Big\}\ .
\end{align}
To lower the number  of derivatives we first replace the terms that are quadratic in $\rho _{\mu \nu } $ by the kinetic term of an
auxiliary field $\chi _{\mu }$, while adding another term with a Lagrange multiplier $\psi_\mu$  to fix the relation between
$\rho _{\mu \nu }$ and $\chi _{\mu }$:
\begin{align}\label{fermlower}
I_{\text{ferm}}^{\text{lin}}(\text{lower}) =\int d^{3}x\,\Big\{ 4\bar{\rho}_{\mu }\gamma^{\mu\nu\rho}\partial_\nu \rho_\rho
     -4\bar{\chi}_\mu \gamma^{\mu\nu\rho}\partial_\nu \chi_\rho
     -8\bar{\psi}_{\mu }\left( \gamma^{\mu\nu\rho}\rho_{\nu\rho} -m\gamma^{\mu\nu}\chi_\nu \right) \Big\} \ .
\end{align}
The equation of motion for $\psi_\mu$ enables us to express $\chi_\mu$ in terms of $\rho_{\mu\nu}$. The result is given in
eq.~\eqref{solchi}. Substituting this solution for $\chi_\mu $ back into the action, the terms linear in the Lagrange multiplier
$\psi_\mu$ drop out and we re-obtain the higher-derivative fermionic action given in eq.~\eqref{fermhigher}.

Adding up the lower-derivative bosonic action \eqref{lowNMG} and the lower-derivative fer\-mio\-nic action \eqref{fermlower}
we obtain the lower-derivative supersymmetric action \eqref{linearizedNMG1}, albeit without the bosonic auxiliary
fields $(M,\,N,\,P,\,A_\mu)$. We only consider a formulation in which these auxiliary fields are absent.

Having introduced the new auxiliary fields $(q_{\mu\nu}\,,\psi_\mu\,,\chi_\mu)$ we should derive their supersymmetry rules.
They can be derived by starting from the solutions \eqref{solqu}, \eqref{solchi} and \eqref{solpsi} of
these auxiliary fields in terms of $h_{\mu\nu}$ and $\rho_\mu$ and applying the supersymmetry rules of $h_{\mu\nu}$ and $\rho_\mu$
given in eq.~\eqref{linearizedNMG4}. This leads to supersymmetry rules that do not contain the auxiliary fields. These can be
introduced by adding to the supersymmetry rules a number of (field-dependent) equation of motion symmetries.
We thus find the intermediate result:
\begin{align}\begin{split}\label{linearizedNMGmod}
\delta h_{\mu \nu } &=\bar{\epsilon}\gamma_{(\mu }\rho_{\nu)} \ , \\[.1truecm]
\delta \rho_\mu &= -\frac14\gamma^{\rho\sigma}\partial_\rho h_{\mu\sigma}\epsilon +\frac12S\gamma_\mu\epsilon \ , \\[.1truecm]
\delta S &= \frac14\bar \epsilon\gamma^{\mu\nu}\rho_{\mu\nu} \,,\\[.1truecm]
\delta q_{\mu \nu } &= \bar{\epsilon}\gamma_{(\mu }\psi_{\nu)}+\frac{1}{m}\bar{\epsilon}\partial_{(\mu }\chi_{\nu )}\ , \\[.1truecm]
\delta \psi_\mu &= -\frac14\gamma^{\rho\sigma}\partial_\rho q_{\mu\sigma} \epsilon \ , \\[.1truecm]
\delta \chi_\mu &=\frac{m}{4}\gamma^\nu q_{\mu\nu}\epsilon + \frac{1}{2m} \epsilon\partial_\mu S  \ .
\end{split}\end{align}
These transformation rules are not yet quite the same as the ones given in eq.~\eqref{linearizedNMG2}.
In particular, the transformation
rules of $S$ and $\chi_\mu$ are different. The difference is yet another  ``on-shell symmetry''
of the action eq.~\eqref{linearizedNMG1},
with spinor parameter $\eta$, given by
\begin{align}\begin{split}
\delta S &=  -\frac14\bar\eta\gamma^{\mu\nu}\psi_{\mu\nu} \ , \\[.1truecm]
\delta \chi_\mu &= -\frac{1}{2m}\eta\partial_\mu S \ .
\end{split}\end{align}
The transformation rules in eqs.~\eqref{linearizedNMG2} and \eqref{linearizedNMGmod} are therefore equivalent up to an on-shell
symmetry with parameter $\eta=\epsilon$:
\begin{align}
\delta_{\text{susy}}(\text{eq.}~\eqref{linearizedNMG2}) =
 \delta_{\text{susy}}(\text{eq.}~\eqref{linearizedNMGmod}) + \delta_{\text{on-shell}}(\eta=\epsilon) \ .
\end{align}

We now wish to discuss in which sense the previous analysis can be
extended to the non-linear case. For simplicity, we take the
approximation in which one considers only the terms in the action
that are independent of the fermions and the terms that are
bilinear in the fermions. Furthermore, we ignore in the
supersymmetry variation of the action terms that depend on the
auxiliary scalar $S$. Since terms linear in $S$ only occur in terms
bilinear in fermions this effectively implies that we may set $S=0$
in the action. In this approximation the higher-derivative action of
SNMG is given by \cite{Andringa:2009yc}
\begin{align}\label{non-linear SNMG action truncated}
I^{\text{nonlin}}_{\text{SNMG}} &\left( \text{higher}\right) =\int d^{3}x\ e\,\Big\{
   -4R\left( \hat{\omega} \right) +\frac{1}{m^{2}}R^{\mu \nu ab}\left( \hat{\omega}\right) R_{\mu \nu ab}\left( \hat{\omega}\right)
   -\frac{1}{2m^{2}}R^{2}\left( \hat{\omega} \right) \nonumber \\[.2truecm]
 & +4 \bar{\rho}_\mu \gamma^{\mu\nu\rho} D_\nu \left( \hat{\omega}\right) \rho_{\rho}
   +\frac{8}{m^{2}}\bar{\rho}_{ab}\left( \hat{\omega}\right) \slashed D\left( \hat{\omega}\right) \rho^{ab}\left( \hat{\omega}\right)
   -\frac{2}{m^{2}}\bar\rho_{\mu\nu}(\hat\omega)\gamma^{\mu\nu}\slashed D(\hat\omega) \gamma^{\rho\sigma}\rho_{\rho\sigma}(\hat\omega)
  \nonumber \\[.2truecm]
 & -\frac{2}{m^{2}}R_{\mu \nu ab}\left(\hat\omega\right) \bar{\rho}_{\rho}\gamma^{\mu\nu}\gamma^\rho \rho^{ab}\left(\hat\omega\right)
  -\frac{2}{m^{2}}R\left(\hat\omega\right) \bar{\rho}^{\mu}\gamma^{\nu}\rho_{\mu\nu}\left( \hat{\omega}\right) \\[.2truecm]
 & +\text{ higher-order fermions and S-dependent terms}\Big\} \ . \nonumber
\end{align}
Note that we have replaced the symmetric tensor $h_{\mu\nu}$ by a
Dreibein field $e_\mu{}^a$. Keeping the same approximation discussed
above the action \eqref{non-linear SNMG action truncated} is
invariant under the supersymmetry rules
\begin{align}\begin{split}\label{susyrules1}
 \delta e_{\mu }{}^{a} &=\frac{1}{2}\bar{\epsilon}\gamma^a\rho_\mu \ , \\[.2truecm]
 \delta \rho_\mu  &= D_{\mu }\left( \hat{\omega}\right) \epsilon \ .
\end{split}\end{align}

We first consider the lowering of the number of derivatives in the bosonic part of the action. Since the Ricci tensor
now depends on a torsion-full spin connection we need a {\sl non-symmetric} auxiliary tensor $q_{\mu,\nu}$. The action
\eqref{non-linear SNMG action truncated} can then be converted into the following equivalent action:
\begin{align}
 I^{\text{nonlin}}_{\text{SNMG}} &\left( \text{higher}\right)= \int d^{3}x\ e\,\Big\{
        -4R\left( \hat\omega\right) -m^{2}\left( q^{\mu,\nu}q_{\mu ,\nu }-q^{2}\right)
        +2q^{\mu ,\nu}G_{\mu, \nu }\left( \hat\omega\right) \notag \\[.2truecm]
  &  +4 \bar{\rho}_\mu \gamma^{\mu\nu\rho} D_\nu \left(\hat\omega\right) \rho_\rho
     +\frac{8}{m^{2}}\bar{\rho}_{ab}\left( \hat\omega\right) \slashed D\left(\hat\omega\right) \rho^{ab}\left(\hat\omega\right)
     -\frac{2}{m^{2}}\bar\rho_{\mu\nu}(\hat\omega)\gamma^{\mu\nu}\slashed D(\hat\omega)\gamma^{\rho\sigma}\rho_{\rho\sigma}(\hat\omega)
  \notag \\[.2truecm]
 &  -\frac{2}{m^{2}}R_{\mu \nu ab}(\hat\omega) \bar\rho_\rho \gamma^{\mu\nu}\gamma^\rho \rho^{ab}(\hat\omega)
   -\frac{2}{m^{2}}R (\hat\omega) \bar\rho^\mu \gamma^\nu \rho_{\mu\nu}\left( \hat{\omega}\right)
  \label{non-linear SNMG action truncated2}\\[.2truecm]
 & +\text{ higher-order fermions and S-dependent terms}\Big\}\ . \notag
\end{align}
The equivalence with the previous action can be seen by solving the equation of motion for $q_{\mu,\nu}$:
\begin{align}\label{solution2}
 q_{\mu ,\nu }=\frac{1}{m^{2}}G_{\mu, \nu }\left( \hat{\omega}\right) -\frac{1 }{2m^{2}}g_{\mu \nu }G^{\text{tr}}\left(
 \hat{\omega}\right)
\end{align}
and substituting this solution back into the action. Note that the solution for $q_{\mu,\nu}$ is not super-covariant.

We next  consider the lowering of the number of derivatives in the fermionic terms in the action. Following the linearized case we
define an auxiliary vector-spinor $\chi_\mu$ as
\begin{align}
 \chi_\mu = -\frac{1}{2m}\gamma^{\rho\sigma} \gamma_\mu \rho_{\rho\sigma}\left(\hat\omega\right) \ ,\label{chi in terms of rho non-linear}
\end{align}
or equivalently
\begin{align}
 \rho_{\mu\nu}\left(\hat\omega\right) = -m\gamma_{[ \mu }\chi_{\nu ]} \ . \label{rho in terms of chi non-linear}
\end{align}
The first equation is the non-linear generalization of eq.~\eqref{solchi}. Using this definition one can show the following identity
\begin{align}\begin{split}\label{identity2}
 \frac{8}{m^{2}}e\bar{\rho}_{ab} & (\hat\omega) \slashed D(\hat\omega) \rho^{ab}(\hat\omega)
    -\frac{2}{m^{2}}e \bar\rho_{\mu\nu}(\hat\omega) \gamma^{\mu\nu}\slashed D(\hat\omega)
               \left[\gamma^{\rho\sigma}\rho_{\rho \sigma}(\hat\omega) \right]  = \\[.2truecm]
 & =-4e\bar{\chi}_\mu \gamma^{\mu\nu\rho}D_\nu (\hat\omega) \chi_\rho
    -\frac{1}{m}eR_{\mu \nu ab}(\hat\omega) \bar{\rho}_\rho \gamma^{\mu\nu\rho}\gamma^{ab}\gamma^\sigma \chi_\sigma \\[.2truecm]
 & \quad +\,\text{higher-order fermions and total derivative terms} \ ,
\end{split}\end{align}
which is the non-linear generalization of the identity \eqref{identity1}.
This identity can be used to replace the higher-derivative kinetic terms of the fermions by lower-derivative ones.
At the same time we may use eq.~\eqref{rho in terms of chi non-linear} to replace $\rho_{\mu\nu}$ by $\chi_\mu$.
This can be done by introducing a Lagrange multiplier $\psi_\mu$ whose equation of motion allows us to use
eq.~\eqref{chi in terms of rho non-linear}. This leads to the following action:
\begin{align}
 I_{\text{SNMG}}^{\text{nonlin}}\left( \text{lower}\right) &=\int d^{3}x\ e\,\Big\{
   -4R\left( \hat{\omega} \right) +2q^{\mu,\nu}G_{\mu,\nu}\left( \hat{\omega}\right)
   -m^{2}\left( q^{\mu ,\nu }q_{\mu ,\nu }-q^{2}\right)  \notag \\[.2truecm]
 & +4\bar{\rho}_{\mu }\gamma^{\mu\nu\rho}D_{\nu }\left( \hat{\omega}\right) \rho_{\rho}
   -4\bar{\chi}_{\mu }\gamma^{\mu\nu\rho}D_{\nu }\left( \hat{\omega}\right) \chi_{\rho}
   -8\bar{\psi}_{\mu } \gamma^{\mu\nu\rho}\rho_{\nu\rho}\left( \hat{\omega}\right)
   +8m\bar{\psi}_{\mu }\gamma^{\mu\nu}\chi_\nu \notag \\[.2truecm]
 & -\frac{1}{m} R_{\mu \nu ab}\left(\hat{\omega}\right) \bar{\rho}_{\rho }\gamma^{\mu\nu\rho}\gamma^{ab}\gamma^\sigma \chi_\sigma
   +\frac{2}{m}R_{\mu \nu ab}\left(\hat{\omega}\right) \bar{\rho}_{\rho}\gamma^{\mu\nu}\gamma^\rho \gamma^a\chi^b  \notag\\[.2truecm]
 & -\frac{1}{m}R\left( \hat{\omega}\right)\bar{\rho}_{\mu }\gamma^{\mu\nu}\chi_\nu
   -\frac{2}{m}R\left( \hat{\omega}\right) \bar{\rho}^{\mu }\chi_\mu  \notag \\[.2truecm]
 & +\,\text{higher-order fermions and S-dependent terms}\Big\} \ .
\end{align}

Our next task is to derive the supersymmetry rules of the  auxiliary fields $q_{\mu,\nu}\,, \psi_\mu$ and $\chi_\mu$. Using
the solutions of the auxiliary fields in terms of $e_\mu{}^a$ and $\rho_\mu$ we derived these supersymmetry rules. In this
way one obtains supersymmetry rules that  do not contain any of the auxiliary fields and, consequently, do not reduce to the
supersymmetry rules \eqref{linearizedNMG2} upon linearization. To achieve this, we must add to these
transformation rules
a number of field-dependent equation of motion symmetries, like we did in the linearized case.
Since the results we obtained are not illuminating we refrain from giving the explicit expressions here.

A disadvantage of the present approach is that, although in principle possible in the approximation we considered, one cannot maintain
the interpretation of $S$ as a torsion contribution to the spin-connection. This makes the result rather cumbersome.
It would be interesting to see whether a superspace approach could improve on this. Without further insight
the lower-derivative formulation of SNMG, if it exists at all at the full non-linear level, does not take the same elegant
form as the higher-derivative formulation presented in \cite{Andringa:2009yc}.

\section{Conclusions}\label{sec:sec6}

In this work we considered the ${\cal N}=1$ supersymmetrization of New Massive Gravity in the presence of auxiliary fields.
All auxiliary fields are needed to close the supersymmetry algebra off-shell. At the linearized level, we distinguished
between two types of auxiliary fields: the ``non-trivial'' ones  whose elimination leads to higher derivatives in the
Lagrangian (these are the fields $q_{\mu\nu}, \psi_\mu$ and $\chi_\mu$) and the ``trivial'' ones whose elimination (if
possible at all at the full non-linear level) does not lead to higher
derivatives (these are the fields $S, M, N, P$ and $A_\mu$). We found that at the linearized level all auxiliary fields could
be included leading to a linearized SNMG theory without higher derivatives. At the non-linear level we gave a partial answer
for the case that only the trivial auxiliary $S$ and the non-trivial auxiliaries $q_{\mu\nu}, \psi_\mu$ and $\chi_\mu$ were
included. To obtain the full non-linear answer one should perhaps make use of superspace techniques. The answer without
the non-trivial auxiliaries and with higher derivatives can be found in \cite{Andringa:2009yc}.

We discussed a 3D supersymmetric analog of the 4D vDVZ discontinuity by taking the massless limit of the supersymmetric FP
model coupled to a supercurrent multiplet. We showed that in the massless limit there is a non-trivial coupling of a scalar
multiplet (containing the scalar mode $\phi$ of the metric) to a current multiplet (containing the trace of the energy-momentum
tensor). This is the natural supersymmetric extension of what happens in the bosonic case and supports the analysis
of \cite{Deser:1977ur}.

As a by-product we found a way to ``boost up'' the derivatives in the spin-3/2 FP equation, see appendix \ref{app:fermions}.
The trick is based upon the observation that, before boosting up the derivatives like in the construction of the NMG model,
one should first combine the equations of motion describing the helicity +3/2 and $-$3/2 states into a single parity-even
equation with one additional derivative.

It is natural to extend the results of this work to the case of extended, i.e.~$\mathcal{N}>1$,  supersymmetry, or to
'cosmological' massive gravity theories. Higher-derivative, linearized versions of NMG with extended supersymmetry, or
anti-de Sitter vacua, were given in \cite{Bergshoeff:2010mf,Bergshoeff:2010ui}.
Of special interest is  the case of maximal supersymmetry since this would correspond to the
KK reduction of the $\mathcal{N}=8$ massless maximal supergravity multiplet which only exists in a formulation
without (trivial) auxiliary fields. We expect that having a formulation of this maximal SNMG
theory without higher-derivatives will be
useful in finding out whether this massive 3D supergravity model has the same miraculous ultraviolet properties
as in the 4D massless case.

\begin{appendix}

\section{Off-shell $\mathcal{N}=1$ Massless Multiplets}\label{app:multiplets}

In this appendix we collect the off-shell formulations of the different 3D massless multiplets with $\mathcal{N}=1$ supersymmetry. A useful reference where more properties about 3D supersymmetry can be found is
\cite{Gates:1983nr}. The field content of the different multiplets can be found in Table \ref{table:1}. \vskip .2truecm

\begin{table}[h]
\begin{center}
{\small
\begin{tabular}{|c|c|c|c|}
\hline\rule[-1mm]{0mm}{6mm}
multiplet&fields&off-shell&on-shell\\[.1truecm]
\hline \rule[-1mm]{0mm}{6mm}
$s=2$&$h_{\mu\nu}\,, \psi_\mu\,, S$&4+4&0+0\\[.05truecm]
\hline \rule[-1mm]{0mm}{6mm}
$s=1$&$V_\mu\,,N\,,A_\mu\,,\chi_\mu\,,\psi$&6+6&1+1\\[.05truecm]
\hline \rule[-1mm]{0mm}{6mm}
$s=0$&$\phi\,,\chi\,, F$&2+2&1+1\\[.05truecm]
\hline \hline\rule[-1mm]{0mm}{6mm}
gravitino multiplet&$\chi_\mu\,, A_\mu\,, D$&4+4&0+0\\[.05truecm]
\hline \rule[-1mm]{0mm}{6mm}
vector multiplet&$V_\mu\,, \psi$&2+2&1+1\\[.05truecm]
\hline
\end{tabular}
}
\end{center}
 \caption{\sl This Table indicates the field content and off-shell/on-shell
 degrees of freedom of the different massless multiplets. Only the massless multiplets above the double horizontal line occur in the massless limit of the FP model.
 }\label{table:1}
\end{table}
\vskip .5truecm

\noindent {\bf s=2}\ \ \ The off-shell version of the 3D massless spin-2 multiplet is well-known. The multiplet is extended
with an auxiliary real scalar field $S$. The off-shell supersymmetry rules are given by
\begin{align}
\delta h_{\mu \nu} &= \bar{\epsilon} \gamma_{(\mu} \psi_{\nu)} \ ,\nonumber \\[.1truecm]
\delta \psi_\mu & = -\frac14 \gamma^{\rho \sigma} \partial_\rho h_{\mu \sigma} \epsilon +\frac12 S\gamma_\mu\epsilon\ ,\\[.1truecm]
\delta S & = \frac14\bar\epsilon \gamma^{\mu\nu}\psi_{\mu\nu}\ ,\nonumber
\end{align}
where
\begin{align}
\psi_{\mu\nu} = \frac12\big(\partial_\mu\psi_\nu -\partial_\nu\psi_\mu\big)\ .
\end{align}
These transformation rules leave the following action  invariant:
\begin{align}\label{s=2action}
I_{s=2} =\int d^{3}x\,\Big\{ h^{\mu \nu }G^{\text{lin}}_{\mu \nu}\left( h\right)
      -4\bar{\psi}_{\mu }\gamma^{\mu\nu\rho}\partial_\nu \psi_\rho -8 S^2\Big\}\ .
\end{align}
\vskip .5truecm

\noindent {\bf s=1}\ \ \ The off-shell ``mixed gravitino-vector'' multiplet consists of a propagating vector $V_\mu$,
an auxiliary vector $A_\mu$, an auxiliary scalar $N$, a vector spinor $\chi_\mu$ and a spinor $\psi$.
An on-shell version of this multiplet, called ``vector-spinor'' multiplet, has been considered in \cite{Nishino:2011zzd}.
The off-shell supersymmetry rules are given by
\begin{align}\label{vectorspinormulitplet}
\delta V_\mu &= \bar\epsilon\gamma_\mu\psi-\frac12\bar\epsilon\chi_\mu \ ,\nonumber\\[.1truecm]
\delta \psi &=-\frac18\gamma^{\rho\lambda}F_{\rho\lambda}\epsilon -\frac{1}{12} N\epsilon
               -\frac{1}{12}\gamma^\alpha A_\alpha\epsilon\ ,\nonumber\\[.1truecm]
\delta \chi_\mu &= -\frac14\gamma^\alpha F_{\alpha\mu}\epsilon -\frac18\gamma_\mu\gamma^{\rho\lambda}F_{\rho\lambda}\epsilon
               -\frac16\gamma_\mu N\epsilon +\frac14A_\mu\epsilon -\frac16\gamma_\mu\gamma^\alpha A_\alpha\epsilon\ ,\\[.2truecm]
\delta N &= \bar\epsilon\gamma^\alpha\partial_\alpha\psi
             -\bar\epsilon\gamma^{\alpha\beta}\partial_\alpha\chi_\beta \ ,\nonumber\\[.1truecm]
\delta A_\mu &= \frac32\bar\epsilon\gamma_\mu^{\phantom{\mu}\alpha\beta}\partial_\alpha\chi_\beta
             -\bar\epsilon\gamma_\mu\gamma^{\alpha\beta}\partial_\alpha\chi_\beta
             +\bar\epsilon\gamma_\mu^{\phantom{\mu}\alpha}\partial_\alpha\psi+\bar\epsilon\partial_\mu\psi \ .\nonumber
\end{align}
Note that this multiplet is irreducible. It cannot be written as the sum of a  gravitino and vector multiplet.
These multiplets are given below. The supersymmetric action for this multiplet is given by
\begin{align}
I_{s=1} = \int d^{3}x\,\Big\{-F^{\mu\nu}F_{\mu\nu}-\frac23N^2+\frac23A_\mu A^\mu
              -4\bar\chi_\mu\gamma^{\mu\nu\rho}\partial_\nu\chi_\rho-8\bar\psi\gamma^\mu\partial_\mu\psi\Big\}\ ,
\end{align}
with $F_{\mu\nu} =\partial_\mu V_\nu -\partial_\nu V_\mu$\,.
\vskip .5truecm

\noindent {\bf s=0}\ \ \ The off-shell scalar multiplet consists of a scalar $\phi$, a spinor $\chi$ and an auxiliary scalar $F$.
The off-shell supersymmetry rules are given by
\begin{align}\label{masslessS}
\delta \phi &= \frac14\bar\epsilon\chi \ ,\nonumber\\[.1truecm]
\delta\chi &=\gamma^\mu\epsilon\,(\partial_\mu\phi) -\frac14 F\epsilon \ ,\\[.2truecm]
\delta F &= -\bar\epsilon \gamma^\mu\partial_\mu\chi \ .\nonumber
\end{align}
The supersymmetric action for a scalar multiplet is given by
\begin{align}\label{masslessactionS}
I_{s=0} = \int d^{3}x\,\Big\{-\partial^\mu\phi\partial_\mu\phi -\frac14\bar\chi\gamma^\mu\partial_\mu\chi +\frac{1}{16}F^2\Big\}\ .
\end{align}
\vskip .2truecm

Besides the massless multiplets discussed so-far there is a separate gravitino and vector multiplet. The vector multiplet arises
in section \ref{sec:sec2} in the massless limit of the Proca theory. For completeness we give these two multiplets below.
\bigskip

\noindent {\bf gravitino multiplet}\ \ \ The off-shell gravitino multiplet consists of a gravitino $\chi_\mu$, an auxiliary
vector $A_\mu$ and an auxiliary scalar $D$. The off-shell supersymmetry rules are given by
\begin{align}
\delta \chi_\mu &= \frac14\gamma^\lambda\gamma_\mu\epsilon A_\lambda + \frac12\gamma_\mu\epsilon D\ ,\nonumber\\[.1truecm]
\delta A_\mu &= \frac12\bar\epsilon\gamma^{\rho\sigma}\gamma_\mu \chi_{\rho\sigma} \ ,\\[.1truecm]
\delta  D &= \frac14\bar\epsilon \gamma^{\rho\sigma}\chi_{\rho\sigma} \ ,\nonumber
\end{align}
where
\begin{align}
 \chi_{\mu\nu} = \frac12\big(\partial_\mu\chi_\nu -\partial_\mu\chi_\mu\big)\ .
\end{align}
These transformation rules leave the following action  invariant:
\begin{align}\label{s=3/2action }
I_{s=3/2} = \int d^3 x\,\Big\{-4\bar{\chi}_\mu \gamma^{\mu\nu\rho}\partial_\nu \chi_\rho -\frac12 A^\mu A_\mu +2D^2\Big\}\ .
\end{align}
\vskip .5truecm

\noindent {\bf vector multiplet}\ \ \ The off-shell vector multiplet consists of a vector $V_\mu$ and a spinor $\psi$. The
off-shell supersymmetry rules are given by
\begin{align}\label{masslessV}
\delta V_\mu &= -\bar\epsilon\gamma_\mu\psi \ ,\nonumber \\[.1truecm]
\delta \psi &= \frac18\gamma^{\mu\nu}\epsilon \, F_{\mu\nu} \ ,
\end{align}
with $F_{\mu\nu} =\partial_\mu V_\nu -\partial_\nu V_\mu$\,. The supersymmetric action for a vector multiplet is given by
\begin{align}\label{masslessactionV}
I_{s=1} = \int d^{3}x\,\Big\{-\frac14 F^{\mu\nu}F_{\mu\nu} -2 \bar\psi\gamma^\mu\partial_\mu\psi\Big\}\ .
\end{align}
\bigskip

This finishes our discussion of the massless multiplets in three dimensions.

\section{Boosting up the Derivatives in Spin-3/2 FP}\label{app:fermions}

In this appendix we show how the higher-derivative kinetic terms for the gravitino $\rho_\mu$ can be obtained by boosting up
the derivatives in the massive spin-3/2 FP equations in the same way as that has been done for the spin-2 FP equations in the
construction of New Massive Gravity  \cite{Bergshoeff:2009hq} except for one subtlety.

Our starting point is the following fermionic action with two massive gravitini, $\psi_\mu$ and $\chi_\mu$, each of which carries
only one physical degree of freedom in 3D,
\begin{align}\label{actB1}
I\left[ \psi ,\chi \right] =\int d^{3}x\,\Big\{ -4\bar{\psi}_\mu \gamma^{\mu\nu\rho}\partial_\nu \psi_\rho
     -4\bar{\chi}_{\mu }\gamma^{\mu\nu\rho}\partial_\nu \chi_\rho +8m\bar{\psi}_\mu \gamma^{\mu\nu}\chi_\nu \Big\} \ .
\end{align}
The equations of motion following from this action are given by
\begin{align}\label{eqB2}
\gamma^{\mu\nu\rho}\partial_\nu \psi_\rho -m\gamma^{\mu\nu}\chi_\nu &=0 \ , \hskip 1truecm
 \gamma ^{\mu\nu\rho}\partial_\nu \chi_\rho -m\gamma^{\mu\nu}\psi_\nu =0 \ .
\end{align}
Note that each one of the equations \eqref{eqB2} can be used to solve for one gravitino in terms of the other one. However,
this solution does not solve the other equation. Therefore, one cannot substitute only one solution back into \eqref{actB1}
because one would lose information about the differential constraint encoded in the other equation.

After diagonalization
\begin{align}
\zeta_\mu^1 =\psi_\mu +\chi_\mu \ , \hskip 1truecm \zeta_\mu^2 =\psi_\mu -\chi_\mu \ .
\end{align}
we obtain the massive FP equations for a helicity +3/2 and -3/2 state:
\begin{align}
\left( \slashed\partial +m\right) \zeta_\mu^1 = 0\ ,\ \gamma^\mu \zeta_\mu^1=0\ ,\ \partial^\mu \zeta_\mu^1=0\ , \\[.2truecm]
\left( \slashed\partial -m\right) \zeta_\mu^2 = 0\ ,\ \gamma^\mu \zeta_\mu^2=0\ ,\ \partial^\mu \zeta_\mu^2=0\ .
\end{align}

To boost up the derivatives in these equations we may proceed in two ways. One option is to boost up the derivatives in each equation
separately by solving the corresponding differential constraint. In a second step one should then combine the two higher-derivative
equations by a single equation in terms of $\rho_\mu$ by a so-called ``soldering'' technique which has also been applied to construct New
Massive Gravity out of two different Topologically Massive Gravities \cite{Dalmazi:2009es}. Alternatively, it is more convenient to first
combine the two equations into the following equivalent second-order equation which is manifestly parity-invariant:
\begin{align}
\left(\Box -m^2\right) \zeta_\mu = \left(\slashed\partial\mp m\right)\left(\slashed\partial \pm m\right)\zeta_\mu = 0
   \ ,\hskip 1truecm \gamma^\mu \zeta_\mu =0\ ,\ \partial^{\mu }\zeta_\mu =0 \ . \label{K-G tr-less div-less}
\end{align}
Note that the action corresponding to these equations of motion cannot be used in a supersymmetric action since the fermionic
kinetic term would have the same number of derivatives as the standard bosonic kinetic term describing a spin-2 state.

We are now ready to perform the procedure of \textquotedblleft boosting up the derivatives\textquotedblright\ in the same way as in the
bosonic theory where it leads to the higher-derivative NMG theory. To be specific, we solve the divergenceless
condition $\partial ^{\mu }\zeta _{\mu }=0$ in terms of a new vector-spinor $\rho_\mu$ as follows:
\begin{align}
\zeta_\mu =\mathcal{R}_\mu \left( \rho \right) \equiv \varepsilon_\mu{}^{\nu \rho }\partial_\nu \rho_\rho \ .
\end{align}
Substituting this solution back into the other two equations in (\ref{K-G tr-less div-less}) leads to the higher-derivative
equations
\begin{align}
\left(\Box -m^2\right) \mathcal{R}_\mu \left( \rho \right) =0\ ,\hskip 1truecm \gamma^\mu \mathcal{R}_\mu \left(\rho\right) =0
  \ . \label{higher derivative K-G tr-less}
\end{align}
These equations of motion are invariant under the gauge symmetry
\begin{align}
\delta \rho_\mu =\partial_\mu \eta \ .
\end{align}
Furthermore, they can be integrated to the following action:
% \footnote{Note that substituting \emph{one} of the solutions of the
% equations \eqref{eqB2} back into the action \eqref{actB1} leads directly to \eqref{actionfinal}. However, this is a mere
% coincidence since it is in general not legitimate to substitute equations of motion back into the action.}
\begin{align}\label{actionfinal}
I\left[ \rho \right] =\int d^{3}x\,\Big\{ \bar{\rho}^{\mu}\mathcal{R}_{\mu }\left( \rho \right)
  -\frac{1}{2m^{2}}\bar{\rho}^{\mu }\slashed\partial \left[\slashed\partial \mathcal{R}_{\mu }\left( \rho \right)
          +\varepsilon_{\mu }{}^{\sigma \tau }\partial_\sigma \mathcal{R}_{\tau }\left(\rho \right) \right] \Big\} \ .
\end{align}
One can show that the equations of motion following from this action implies the algebraic constraint given in
\eqref{higher derivative K-G tr-less}. The action \eqref{actionfinal} is precisely the fermionic part of the action
\eqref{linearizedNMG3} of linearized SNMG.

\end{appendix}

\section*{Acknowledgements}
We thank Olaf Hohm and Paul Townsend for comments on the draft. Y.Y.~would like to thank Rakibur Rahman for inspiring
discussions. Two of us (E.B. and J.R.) would like to thank the  Simons Center for
Geometry and Physics for its hospitality and generous financial support. They wish to thank the
organizers of the Summer Workshop 2012 for providing a stimulating scientific environment in which part of this work
was completed. The work of J.R.~was supported by the START project Y 435-N16 of the Austrian Science Fund (FWF).
L.P.~acknowledges support by the Consejo Nacional de Ciencia y Tecnolog\'ia (CONACyT), the Universidad Nacional
Aut\'onoma de M\'exico via the project UNAM-PAPIIT IN109013 and an Ubbo Emmius sandwich scholarship from the University of
Groningen.
The work of M.K.~and Y.Y.~is supported by the Ubbo Emmius Programme administered by the Graduate School of Science, University
of Groningen. T.Z.~acknowledges support by a grant of the Dutch Academy of Sciences (KNAW).

% \bibliographystyle{utphys}
% \bibliography{massivesusy}

\providecommand{\href}[2]{#2}\begingroup\raggedright\endgroup

\end{document}